\documentclass[9pt,arxiv]{lapreprint}

\usepackage[version=4]{mhchem} 
\usepackage{siunitx}    
\usepackage{pdflscape}  
\usepackage{rotating}   
\usepackage{textgreek}  
\usepackage{gensymb}    
\usepackage[misc]{ifsym} 
\usepackage{orcidlink}  
\usepackage{listings}   
\usepackage{colortbl}   
\usepackage{tabularx}   
\usepackage{longtable}  
\usepackage{subcaption}
\usepackage{multirow}
\usepackage{snotez}     
\usepackage{csquotes}   
\usepackage{prettyref}
\usepackage{amsmath}
\usepackage{todonotes}  
\usepackage[normalem]{ulem}       

\newrefformat{cap}{\hyperref[#1]{Figure~\ref{#1}}}
\newrefformat{fig}{\hyperref[#1]{Figure~\ref{#1}}}
\newrefformat{tab}{\hyperref[#1]{Table~\ref{#1}}}
\newrefformat{sec}{\nameref[#1]{Section~\ref{#1}}}
\newrefformat{sub}{\hyperref[#1]{Section~\ref{#1}}}
\newrefformat{subsec}{\hyperref[#1]{Section~\ref{#1}}}
\newrefformat{cha}{\hyperref[#1]{Chapter~\ref{#1}}}
\newrefformat{lis}{\hyperref[#1]{Listing~\ref{#1}}}
\newrefformat{eq}{\hyperref[#1]{Eq.~\ref{#1}}}

\newcommand{\code}[1]{\texttt{#1}}
\newcommand{\doi}[1]{\href{https://doi.org/#1}{doi:#1}}

\DeclareSIUnit\Molar{M}

\usepackage[			
	backend=biber,      
	bibstyle=apa,
	citestyle=authoryear,
	uniquelist=false,
	natbib=true,		
	hyperref=true,	    
	alldates=year,      
	isbn=false,
	eprint=false,
	doi=true,
	maxcitenames=2,
]{biblatex}

\title{Comparing apples to apples -\\ Using a modular and adaptable analysis pipeline to compare slow cerebral rhythms across heterogeneous datasets}

\author[ \orcidlink{0000-0001-7373-5962} 1,2 \Letter]{Robin Gutzen}
\author[ \orcidlink{0000-0001-7079-5724} 3 ]{Giulia De Bonis}
\author[ \orcidlink{0000-0003-3488-0088} 3,4 ]{Chiara De Luca}
\author[ \orcidlink{0000-0003-0682-1232} 3 ]{Elena Pastorelli}
\author[ \orcidlink{0000-0002-9958-2551} 3 ]{Cristiano Capone}
\author[ \orcidlink{0000-0002-8489-0076} 5,6 ]{Anna Letizia Allegra Mascaro}
\author[ \orcidlink{0000-0002-9605-5852} 5,7 ]{Francesco Resta}
\author[ \orcidlink{0000-0002-8306-0759} 8 ]{Arnau Manasanch}
\author[ \orcidlink{0000-0002-0675-3981} 5,7,9]{Francesco Saverio Pavone}
\author[ \orcidlink{0000-0002-8437-9083} 8,10 ]{Maria V. Sanchez-Vives}
\author[ \orcidlink{0000-0002-2356-4509} 11 ]{Maurizio Mattia}
\author[ \orcidlink{0000-0003-2829-2220} 1,2 ]{Sonja Grün}
\author[ \orcidlink{0000-0003-1937-6086} 3 ]{Pier Stanislao Paolucci}
\author[ \orcidlink{0000-0003-1255-7300} 1 ]{Michael Denker}

\affil[1]{Institute of Neuroscience and Medicine (INM-6) and Institute for Advanced Simulation (IAS-6) and JARA-Institute Brain Structure-Function Relationships (INM-10), Jülich Research Centre, Jülich, Germany}
\affil[2]{Theoretical Systems Neurobiology, RWTH Aachen University, Aachen, Germany}
\affil[3]{Istituto Nazionale di Fisica Nucleare (INFN), Sezione di Roma, Rome, Italy}
\affil[4]{Ph.D.~Program in Behavioural Neuroscience, “Sapienza” University of Rome, Rome, Italy}
\affil[5]{European Laboratory for Non-linear Spectroscopy (LENS), University of Florence, Florence, Italy}
\affil[6]{Neuroscience Institute, National Research Council, Pisa, Italy}
\affil[7]{Department of Physics and Astronomy, University of Florence, Florence, Italy}
\affil[8]{Institut d'Investigacions Biom\`{e}diques August Pi i Sunyer (IDIBAPS), Barcelona, Spain}
\affil[9]{National Institute of Optics, National research Council, Sesto Fiorentino, Italy}
\affil[10]{Instituci\'{o} Catalana de Recerca i Estudis Avan\c{c}ats (ICREA), Barcelona, Spain}
\affil[11]{Natl.~Center for Radiation Protection and Computational Physics, Istituto Superiore di Sanità (ISS), Rome, Italy}

\correspondence{r.gutzen@fz-juelich.de}{}

\presentaddress[]{cerebral cortex, cortical oscillations, slow wave activity, data analysis, software tools, workflow, reproducibility, validation, wide-field calcium imaging, ECoG}

\data[]{Data and code availability is stated in the key resources table in \prettyref{subsec:key_resource_table}. Any additional information required to reanalyze the data reported in this paper is available from the lead contact upon request.}
\compint[]{The authors declare no competing interests.}

\leadauthor{Gutzen}
\shorttitle{Comparing apples to apples}

\begin{document}
\maketitle
\begin{abstract}

Neuroscience is moving towards a more integrative discipline, where understanding brain function requires consolidating the accumulated evidence seen across experiments, species, and measurement techniques.
A remaining challenge on that path is integrating such heterogeneous data into analysis workflows such that consistent and comparable conclusions can be distilled as an experimental basis for models and theories. 
Here, we propose a solution in the context of slow wave activity ($<1$~Hz), which occurs during unconscious brain states like sleep and general anesthesia, and is observed across diverse experimental approaches. 
We address the issue of integrating and comparing heterogeneous data by conceptualizing a general pipeline design that is adaptable to a variety of inputs and applications.
Furthermore, we present the Collaborative Brain Wave Analysis Pipeline (Cobrawap) as a concrete, reusable software implementation to perform broad, detailed, and rigorous comparisons of slow wave characteristics across multiple, openly available ECoG and calcium imaging datasets.

\end{abstract}

\section{Introduction} 
\label{sec:intro}
Today's research landscape excels in an unprecedented richness of experimental data and methodologies. In neurophysiology, this enables an array of research applications and a better calibration of models of brain dynamics and function. However, recording techniques differ considerably in the way in which they capture neural activity. These differences \citep{Hong2019_330, Sejnowski2014_1440} include the type of signal (e.g., electric activity, magnetic fields, fluctuations of calcium concentrations, or radiant isotopes) and the scales on which the signal is observed in terms of temporal resolution (sub-milliseconds to seconds), spatial resolution (micrometer to centimeters), and spatial extent (single electrode to the whole brain). The complementary nature of the different experimental approaches, each one focusing on specific aspects, enables a deeper, comprehensive understanding of neuronal activity. As each recording technique comes with certain trade-offs (e.g., spatial vs.\ temporal resolution, latency, artifacts), it is further possible to validate findings independently of a particular measurement type or device \citep{AllegraMascaro2015_061105}. This raises the challenge to integrate the multi-scale, multi-methodology nature of the data by defining levels of description and relationships between data modalities.

Cross-domain comparisons between different modalities form the foundation of validation scenarios for theories, models, and experimental data to quantify biological variability and obtain a more generalized description of phenomena \citep{Trensch2018_81, Gutzen2018_90, AllegraMascaro2020_31}. However, performing such comparisons is not a trivial task and only rarely addressed. Even if authors adopt definitions and methods from other publications, comparing quantitative findings is not necessarily straightforward. For example, in the studies \citet{Massimini2004_1160} and \citet{Botella-Soler2012_} the same methodology and wave definition is adopted. Still, a direct quantitative relationship between the reported wave velocities ($2.7 \pm 0.2$ m/s and $1.0 \pm 0.2$ m/s, respectively) is difficult due to remaining critical, potentially undocumented differences in the analysis implementations. These make it impossible to accurately retrace the source of the discrepancy, especially when the respective analysis code or data is not accessible or reusable. Due to this uncertainty, it is increasingly ambitious for scientists to interpret and understand the differences in the quantitative results.

The main challenge in such cross-domain comparisons is to find a common basis for the analysis. What this constitutes depends on the involved data types and the scientific questions. More similar data can have more immediate commonalities, whereas very different data may only be compared on an abstract level. Generally, the comparison of two datasets benefits from having a common level of description for the observations, equivalent or at least comparable methods for processing and analyzing, and the use of equivalent implementations and standard algorithms. Indeed, much care is required to eliminate as many potential confounds as possible, as it has been shown that even seemingly feeble influences such as floating point precision, the choice of the operating system, or software versions can have crucial effects on numerical results and add sources of systematic errors \citep{Glatard2015_12}.

A prerequisite of comparability is reproducibility. Any analysis result must first be reliably reproducible with the same data before it can be reasonably compared with results from other data.
Ideally, the analysis results to be compared are generated from the same code base. However, for heterogeneous data, this requires a considerable degree of generality and reusability of the code. A lack of reusability, especially in custom-written code, is a well-known problem and apparent to anyone who ever tried to efficiently build upon the code collection of a former colleague or even their own code from two years ago. The extra effort in creating reusable code is rewarded by being able to bridge otherwise specific and isolated studies and contributing to a more collaborative scientific tool base. Fortunately, many aspects of analysis workflows are already formalized and addressed by open-source software tools and standards, such as data- and metadata representation \citep{Zehl2016_26, Sprenger2019_62, Rubel2021_2021.03.13.435173}, provenance \citep{Butt2020_431}, version control \citep{Bell2017_13}, standardized algorithms and frameworks \citep{Virtanen2020_261, Denker2018_, Omar2014_524}, and workflow management \citep{Molder2021_33, Crusoe2021_, Garijo2017_271}.

A scenario where a multitude of analysis methods exists, acting on observations from a variety of different measurement techniques, spatio-temporal scales and species, is the study of spatially organized activity, in particular wave activity. While waves are a ubiquitous phenomenon their functional roles are not fully understood \citep{Wu2008_487}. 

\begin{figure}
  \begin{center}
    \includegraphics[width=.85\textwidth]{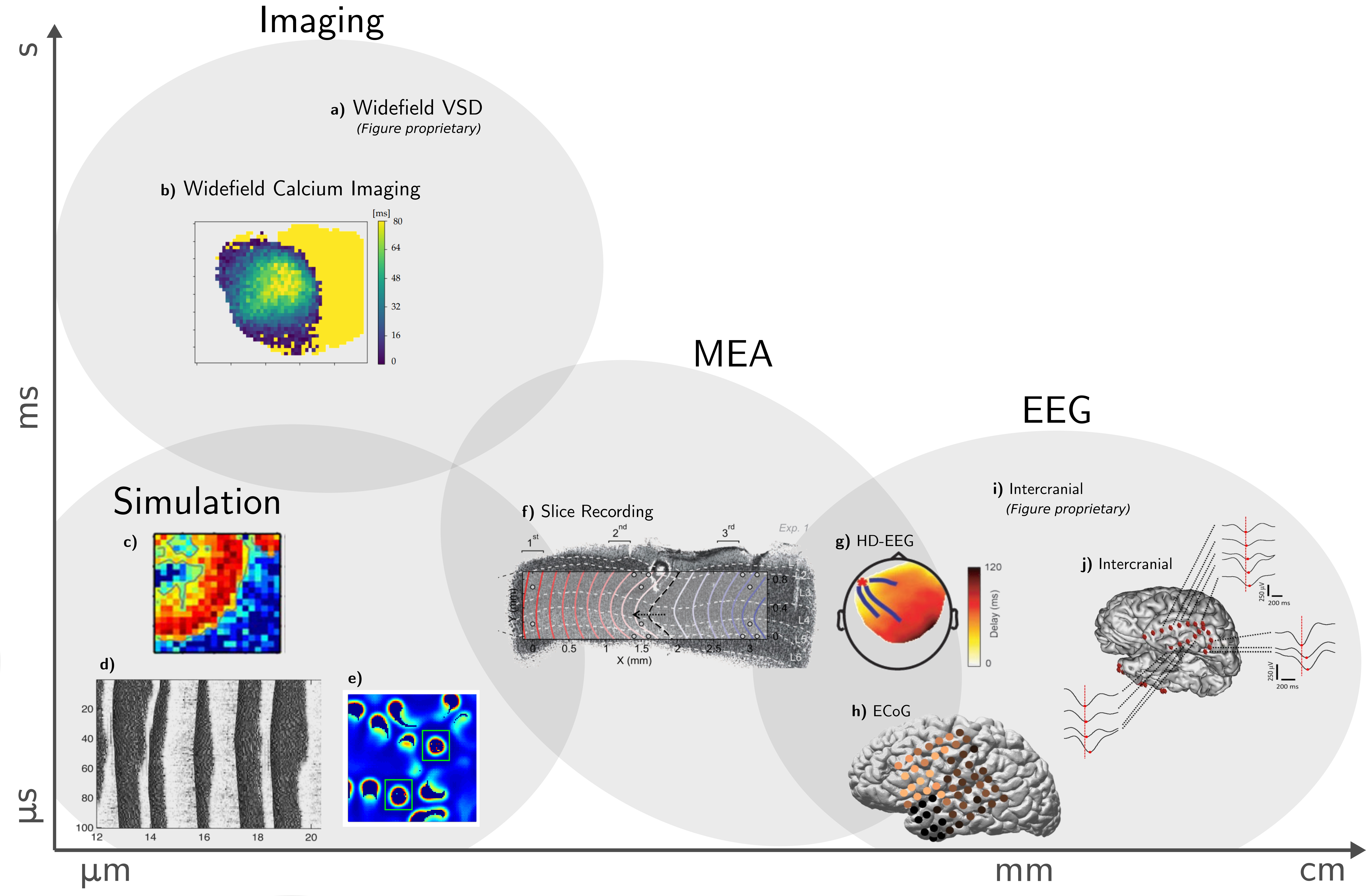}
  \end{center}
  \caption{\textbf{Multiscale, Uniphenomenon: the many faces of slow waves.} \textbf{a)} wide-field voltage-sensitive dye imaging of awake mice \citep{Chan2015_1}, \textbf{b)} recorded anesthetized GCaMP6f mice with wide-field fluorescence microscopy \citep{Celotto2020_14}, \textbf{c}) distributed network of cortical columns of LIF with Spike Frequency Adaptation neurons \citep{Pastorelli2019_33}, \textbf{d)} one-dimensional multi-layer thalamo-cortical model with one- and two-compartment neuron models using Hodgkin-Huxley kinetics \citep{Bazhenov2002_8691}, \textbf{e)} 2D balanced conductance-based spiking neural network model \citep{Keane2015_1591}, \textbf{f)} multi-electrode recording in ferret cortical slices \citep{Capone2019_319}, \textbf{g)} human HD-EEG during first sleep episode of the night \citep{Massimini2004_1160}, \textbf{h)} human ECoG recording during sleep \citep{Muller2016_e17267}, \textbf{i)} intracranial depth EEG in sleeping human subjects \citep{Nir2011_153}, \textbf{j)} intracranial depth EEG in humans during sleep \citep{Botella-Soler2012_}.
  }
  \label{fig:multiscale-uniphenomenon}
\end{figure}

A particular type of wave-like activity that we set out to investigate in this study are slow waves ($< 1$~Hz) \citep{Steriade1993_3252}. They describe propagating activity patterns in the delta band, defined by transitions between states of low activity (Down) and high activity (Up). They are reliably observed in mammals during deep-sleep and anesthesia (\prettyref{fig:multiscale-uniphenomenon}), and are frequently investigated in the study of memory, consciousness, and the cognitive effects of sleep \citep{Hanlon2009_719, Capone2019_1, Golosio2021_e1009045, Staresina2023_2023.01.08.523138}.
For slow waves, findings include that the transitions between Up and Down states are coordinated precisely over a wide cortical range implying a larger network mechanism \citep{Volgushev2006_5665}. The transitions coordinate with the synchronization of the astrocytic network \citep{Szabo2017_6018}, with thalamic activity \citep{Steriade2003_563, Stroh2013_1136, Sheroziya2014_8875}, and across the cortex as reoccurring slow wave patterns can appear over an entire hemisphere \citep{Muller2016_e17267}. Evidence from slice and in-vivo recordings further suggests that wave propagation is guided by excitability, i.e., predominantly resides in layers 4 and 5 \citep{Capone2019_319, Bharioke2022_2024}, and shows distinctly different oscillation characteristics across cortical regions \citep{Ruiz-Mejias2011_2910, Matsui2016_6556, DeBonis2019_70}. Although slow wave activity is characteristic of sleep and anesthesia, it can even be observed in localized areas during wakefulness in EEG recordings of behaving mice \citep{Vyazovskiy2011_443}. Additionally, modeling approaches suggest the importance of long-range connections \citep{Compte2003_2707, Pastorelli2019_33}, synchronous high-amplitude events \citep{Jercog2017_e22425}, and the correct E-I ratio \citep{Compte2009_17, Keane2015_1591} to exhibit propagating slow waves. 
%
In face of such a prevalent phenomenon as slow waves, it is not surprising that the literature reveals a very heterogeneous mosaic of approaches, methods, metrics, and terminology. Due to this plurality, the relationships between the respective findings are rarely apparent and mostly qualitative, limiting the potential of cumulative discovery by the collection of studies. Moreover, it is generally unclear which observables are relevant for the local cortical function or higher cognitive functions (e.g., memory consolidation). The typically reported properties are thus often heuristic and include, for example, transition slopes \citep{Ruiz-Mejias2011_2910}, phase velocity \citep{Massimini2004_1160, Muller2016_e17267}, wave type classification \citep{Townsend2015_4657, Denker2018_1, Roberts2019_1056, Camassa2021_198, Pazienti2022_103918}, source/sink location and propagation patterns \citep{Huang2010_978, Liang2021_3665}, excitability \citep{Mattia2012_239, Ruiz-Mejias2016_3648, DeBonis2019_70}, and event frequency \citep{Capone2022_a}. Thus, we here focus on common observables that can be extracted from different measurement modalities (i.e., planarity, inter-wave intervals, velocity, and direction).
By investigating the relations of these characteristics with parameters such as brain state, anesthetic level, spatial/temporal resolution, etc., we can evaluate the capabilities of measurement techniques, identify biases, constrain theories, develop and benchmark analysis methods, contribute to defining standards, as well as aid the assessment of clinical data of, for example, in the case of coma patients.

In the following, we report two results: first, we conceptually address the problem to develop a flexible formalized approach for constructing analysis pipelines (\prettyref{subsec:results_general_concept}) that we then leverage to implement the Collaborative Brain Wave Analysis Pipeline (Cobrawap) as a modular pipeline to analyze cortical slow wave activity (\prettyref{subsec:results_exemplary_realization}); second, we employ Cobrawap to perform a structured and partially automatized analysis of multiple heterogeneous datasets (\prettyref{subsec:results_dataset_comparison}), and to benchmark the Up state detection method by interchanging the corresponding method block (\prettyref{subsec:results_method_comparison}).

\section{Results}
\label{sec:results}

\subsection{A modular analysis approach enables flexibility in studying slow waves}
\label{subsec:results_general_concept}

\begin{figure}
  \begin{center}
    \includegraphics[width=.85\textwidth]{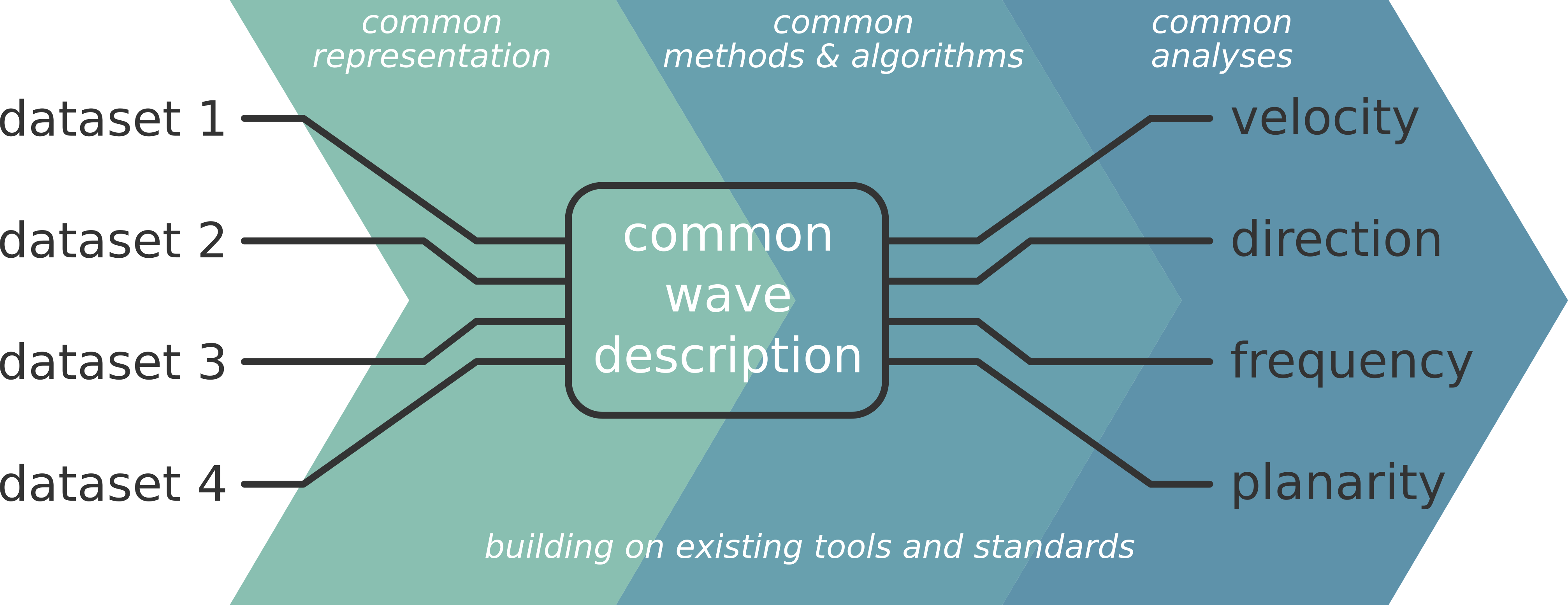}
  \end{center}
  \caption{\textbf{Pipeline Approach.} The proposed pipeline design has the role of aligning methods to create and operate on a common description of the phenomenon of interest. By integrating data from heterogeneous sources on the input side and extracting a variety of common output metrics on the output side, this pipeline approach forms a basis for rigorous comparisons. The pipeline is built on existing tools and standards, e.g., data and metadata representation, file formats, standard packages and implementations, environment handling, and workflow management. The catalog of applicable methods is flexibly extendable, making the analysis pipeline adaptable and reusable.
  }
  \label{fig:pipeline_approach}
\end{figure}

Since there is no single fully comprehensive measure to characterize spatial activity patterns, we focus on identifying commonly used analysis metrics of slow wave activity that enable a comparison between datasets of different measurement types. In designing the pipeline, we first align the heterogeneous input data (e.g. from EEG, implanted electrode arrays, imaging techniques, or simulations) and find a common representation. Although the input data may differ in terms of spatial or temporal resolution, scale, or signal type, we aim to process them by common methods and to converge towards a common description of slow wave activity. From this common description, we derive characterization metrics that are agnostic about the data's origin. In this way, we arrive at comparable slow wave measures and avoid mixing apples and oranges (\prettyref{fig:pipeline_approach}).

The key to making the pipeline adaptable to the different data processing requirements, analysis approaches, and scientific questions is to identify the correct level of modularity. Thus, we first segment the analysis procedure into a series of sequential \textit{stages}. Each stage is a self-consistent logical step in an analysis workflow with a well-defined purpose, input, and output. A stage should be constructed general enough to be reusable as a standalone or in other pipelines. Along the pipeline, subsequent stages become necessarily more specific and tailored towards the scientific application, while the early stages cope with more general tasks such as data integration and preprocessing that are likely shared across different pipelines.

Each stage is further segmented into \textit{blocks}. A block defines a concrete action to be performed on the data, implementing a method. Similar to stages, blocks have a well-defined input and output by which they can be chained together. In contrast to stages, blocks are not necessarily executed in a predefined sequence. Rather, each stage implements the mechanics of the block interactions and defines which block combinations and sequences can be chosen. Some blocks may need to be mandatory for the realization of the stage purpose and have a fixed place in the execution order. Others may be optional and flexibly combined. We identified three basic arrangements to be employed in the stages: \textit{fixed} (a fixed execution order of the blocks), \textit{choose any} (a custom user-defined execution order of a set of blocks), and \textit{choose one} (a multiple-choice selection between blocks for one execution position).

\subsection{Implementation of the Collaborative Brain Wave Analysis Pipeline (Cobrawap)}
\label{subsec:results_exemplary_realization}
Based on the conceptual framework illustrated above (\prettyref{subsec:results_general_concept}), we construct Cobrawap as a specific pipeline application for the analysis and comparison of slow wave activity across 5 publicly available datasets from the EBRAINS Knowledge Graph platform\footnote{\url{https://search.kg.ebrains.eu}} of electrocorticography (ECoG) and wide-field calcium imaging recordings of anesthetized mice. Besides the measurement technique, the 60 examined recordings vary in a range of factors such as experimental setup, the genetic strain of the mice, anesthetic type, anesthesia level, temporal and spatial resolution, and recording duration (see \prettyref{subsec:methods_datasets}).

The pipeline implementation uses the open-source language Python to ensure accessibility and reproducibility. Further, we base the pipeline's architecture on the Python-based workflow manager \textit{snakemake} \citep{Molder2021_33}, which employs input-to-output rules containing executable shell commands (e.g., Python scripts or bash commands). Snakemake structures the execution of the rules by building dependency trees from the final result file(s) back to the initial input, matching the input requirements to the outputs of preceding rules (see \prettyref{subsec:methods_pipeline}).

\begin{figure}
  \begin{fullwidth}
    \begin{center}
      \includegraphics[width=1.1\textwidth]{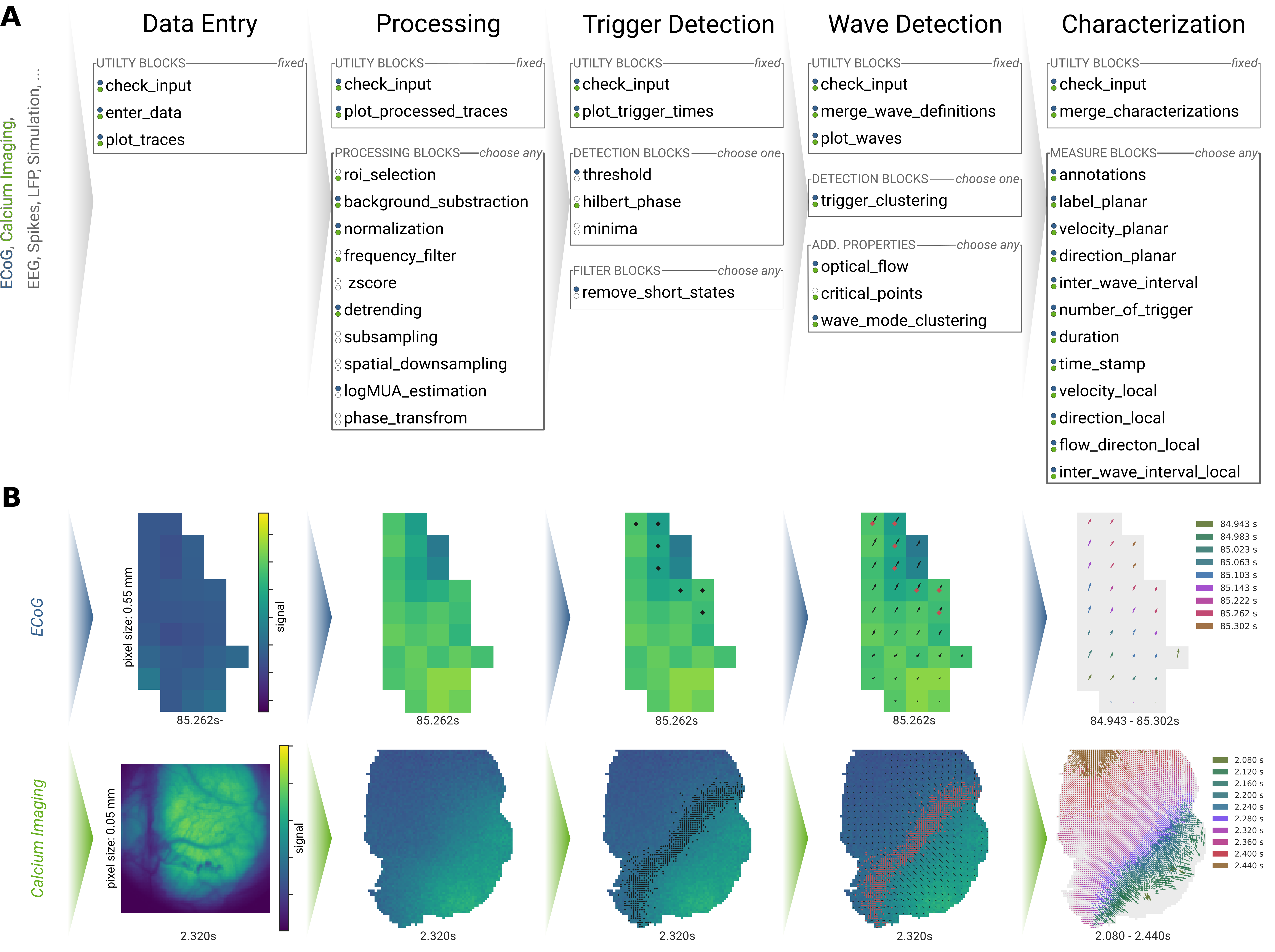}
    \end{center}
    \caption{\textbf{Progression of two datasets (\textit{ECoG} in blue, \textit{wide-field calcium imaging} in green) through the Cobrawap pipeline.} \textbf{A:} The five successive stages contain collections of modular blocks in three different selection modes (fixed, choose one, choose any). The analysis path is adaptable for specific datasets and analyses by selecting and configuring the desired blocks (indicated by colored dots for the datasets). \textbf{B:} The intermediate results after each stage are visualized for the two datasets as color-coded signals on the electrode/pixel grid covering most of the right hemisphere of the mouse brain (up=anterior, right=lateral; ECoG: $4.95\times2.75$~mm, calcium imaging: $5\times5$~mm). From left to right: raw data, post-processed signal, detected upward transitions (black markers), grouped wavefronts (red markers) with the optical flow (arrows), and quantification of the linear flow alignment within the waves (i.e., planarity).}
    \label{fig:pipeline_illustration}
  \end{fullwidth}
  \end{figure}

We organized the Cobrawap pipeline into 5 sequential stages, successively transforming the raw data and extracting slow wave characterizations, as illustrated in \prettyref{fig:pipeline_illustration}. In the following, we describe the role of each stage in the analysis of the ECoG and wide-field calcium imaging data. The stages and blocks are described in detail in the Methods (\prettyref{subsec:methods_pipeline}) and in the corresponding README files.
\begin{itemize}
\item In the first stage, \textbf{Data Entry}, the data is being prepared for the later stages by loading, structuring, and annotating the data and metadata according to the defined representation scheme using the Neo data representation \citep{Garcia2014_10}. For loading the data and converting the highly specific structure of input data into a common representation for further processing by the pipeline, typically, each data source requires a custom script that can be adapted from a template script, making use of Neo for loading and processing data from a variety of file formats and structuring the data. 
\item The second stage, \textbf{Processing}, offers a series of blocks implementing basic pre-processing steps that can be arbitrarily combined. Both data types undergo a background subtraction, a normalization, and a detrending step to remove potential recording artifacts. Considering the different measurement modalities and temporal resolutions of the data types, additionally, the calcium imaging recordings are cut to a region-of-interest and filtered from $0.1$ to $5$~Hz, while the raw ECoG signals are transformed to a logMUA signal (see \prettyref{subsec:logmua_estimation}) with a reduced sampling rate better suited to later capture the slow oscillations and the transitions between Down and Up states.
\item The third stage, \textbf{Trigger Detection}, provides multiple options for a trigger detection method, identifying the time stamps of state transitions (upward or downward trigger) in each channel as an indicator for the possible passage of a wavefront (see \prettyref{subsec:methods_trigger_detection}). In the following, only upward transitions are considered as triggers. Since the logMUA signal shows sharp state transitions, they are best detected by a threshold determined from a channel-wise fit of the amplitude distributions \citep[as in][]{DeBonis2019_70}. Conversely, the transitions in the imaging data are determined by the slow activation function of the fluorescent indicators. Therefore, they are better detected by identifying the upwards slopes by either the Hilbert phase signal crossing a specific value (here $-\pi/2$) or by the local minima preceding a dominant peak. In the following, we use the trigger detection via the Hilbert phase, however, we compare the two methods in \prettyref{subsec:results_method_comparison}.
\item In the fourth stage \textbf{Wave Detection}, the channel-wise trigger times are grouped to define the individual waves (see \prettyref{subsec:trigger_clustering}). This wave representation as a collection of local upward transition times is optionally enriched with additional descriptions such as the optical flow \prettyref{subsec:methods_optical_flow} and the critical points of the resulting vector field \citep{Townsend2018_e1006643} or an additional clustering of the waves into modes, based on the spatial arrangement of the trigger delays.
\item The fifth stage, \textbf{Wave Characterization}, applies a series of quantitative characterizations on the basis of the measures and groupings generated by the previous stages. The selection of characteristics can be tailored toward addressing specific scientific questions or research objectives. To have a consistent output format for this stage, there are two distinct realizations for the fifth stage: one for a characterization using wave-wise measures, e.g., determining one velocity value per wave; and another for a characterization using channel-wise measures, e.g., calculating local velocity values per channel and wave. For simplicity, these two alternatives are presented as a single stage in \prettyref{fig:pipeline_illustration}A.
\end{itemize}

To demonstrate the capabilities of the pipeline approach to generate meaningful quantification of slow wave phenomena, we choose four metrics as the basis for dataset comparisons across datasets: the local (i.e., channel-wise) inter-wave interval, velocity, and direction measures; and the global (i.e., wave-wise) planarity measure. The \textbf{inter-wave interval} is defined as the time delay between the occurrence of two consecutive waves at a recording site. The channel-wise \textbf{velocity} $v$ is calculated from the derivatives of the delay function of a wave $T(x,y)$, which indicates when a wave has reached the position $(x,y)$ in its propagation \citep{Greenberg2018_328, Capone2019_319, Pazienti2022_103918}:
 \begin{equation}
     v_{x,y} = \sqrt{\frac{1}{\partial_x T^2 + \partial_y T^2}}
 \end{equation}
The channel-wise \textbf{direction} of wave propagation can also be derived from the time delay function $T(x,y)$. However, in the following, we use the optical flow of the phase signal (see \prettyref{subsec:methods_optical_flow}). The optical flow is a continuous vector-valued signal for each position $(x,y)$, indicating in which directions the contour lines of equal phase propagate. We define the channel-wise wave directions of a propagating wave as the optical flow vector directions at the time and position of its trigger events. The \textbf{planarity} $P$ of a wave is also defined via the optical flow as the absolute value of the normalized channel-wise direction vectors at the times and positions of all trigger events that belong to a wave, quantifying their alignment on a scale from $0$ to $1$: 
\begin{equation}
    P = \frac{||\sum \vec{v_i}||}{\sum ||\vec{v_i}||}
\end{equation}

\begin{figure}
  \begin{center}
    \includegraphics[width=.95\textwidth]{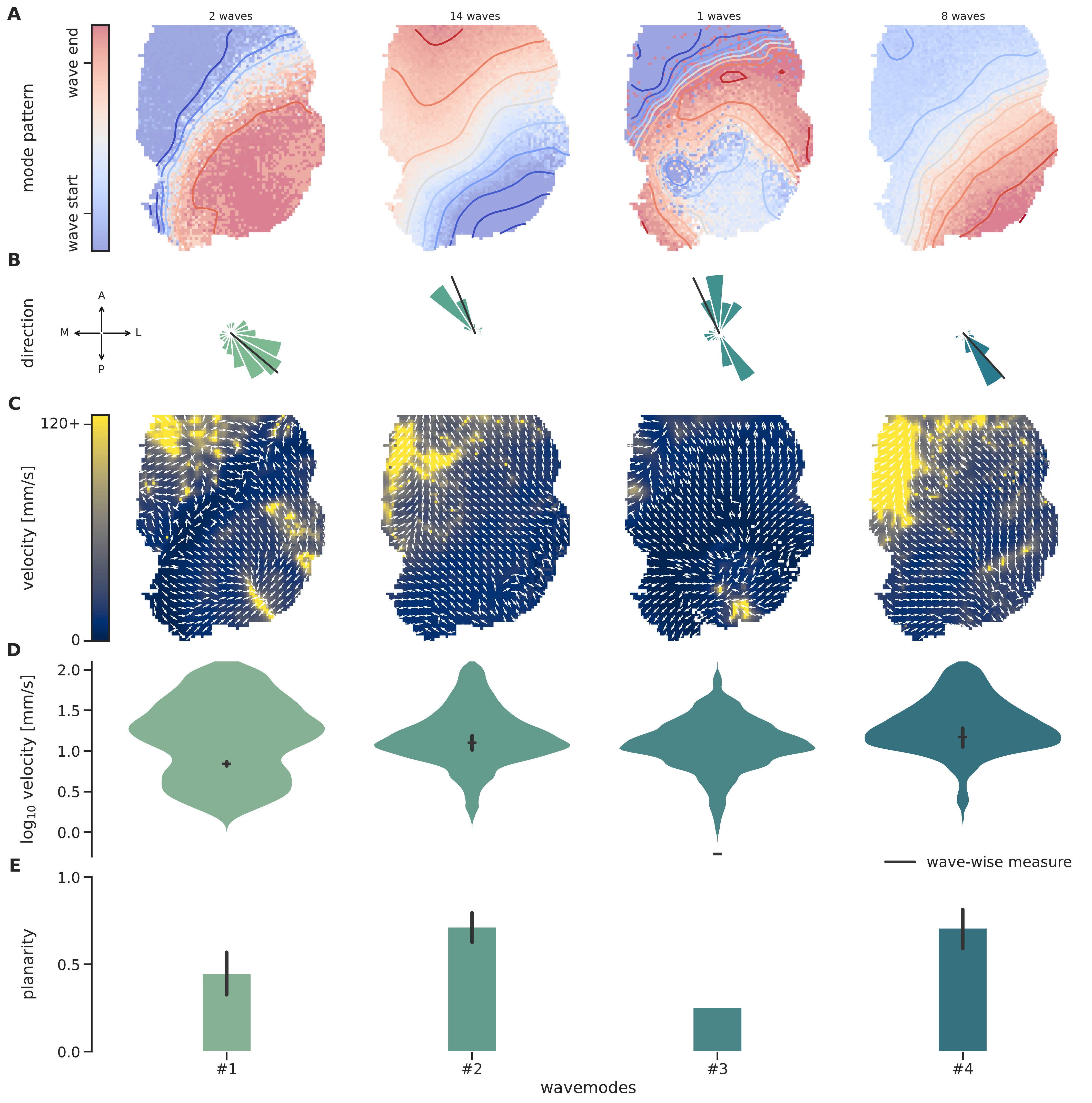}
  \end{center}
  \caption{\textbf{Representation of the Cobrawap pipeline output for one exemplary wide-field calcium imaging recording.} The waves are categorized into one of 4 modes (block wave\_mode\_clustering), shown in columns. \textbf{A:} The average wave pattern of each mode is illustrated as a time-delay heatmap with iso-delay contours. Top: number of waves contributing to the mode. \textbf{B:} The aggregated histograms of channel-wise directions during waves of each mode. The black lines indicate the average wave-wise direction measures. \textbf{C:} Map of the average channel-wise velocities $v_{x,y}$ in waves of each mode (color code), overlayed (arrows) with the average channel-wise direction determined via the optical flow. \textbf{D:} The distributions of channel-wise velocities corresponding to panel C. Black ticks and errorbars: the average and $95\%$ CI of the corresponding wave-wise velocities. \textbf{E:} The average and $95\%$ CI of the planarity $P$ for waves of each mode.}
  \label{fig:wavemodes_dashboard}
\end{figure}

The pipeline output is a table of the characteristic measures derived from the detected wave activity. Supplemental \prettyref{fig:wave_videos} shows videos of the wave activity for two example recordings. \prettyref{fig:wavemodes_dashboard} presents some of the pipeline output measures for one of the calcium imaging recordings. An analogous example figure for an ECoG recording is shown in Supplemental \prettyref{fig:wavemodes_dashboard_ecog}. The channel-wise and wave-wise measured direction and velocity, as well as the wave-wise planarity, are summarized for 4 wave modes, i.e., groups of similar waves, i.e., "wave modes". The wave-mode clustering method (implemented as an optional block in stage 4 of the pipeline) applies a k-means clustering on the trigger delay matrix containing the relative trigger times for each channel in each wave \citep{Ruiz-Mejias2011_2910, Capone2019_319, Pazienti2022_103918}. The number of modes was set by hand to reasonably represent the variability of wave types in the recording. Generally, the 'optimal' number of modes to set for the k-means algorithm depends on recording and the analysis application.

For the presented recording, most waves are relatively planar and travel along the lateral-posterior-to-medial-anterior axis (modes \#2 and \#4). Mode \#1 is a variation of mode \#4 with a lower average velocity, and mode \#3 contains only one wave. 
Although the channel-wise and wave-wise measures for the direction and velocity (\prettyref{fig:wavemodes_dashboard}B,D) are defined and calculated in a different way, they agree considerably well for the modes \#1, \#2, and \#4 when the wave pattern is predominantly planar. The complex wave pattern of mode \#3 cannot be accurately captured by a single wave-wise value for the direction and velocity, resulting in the mismatch with the channel-wise measures. Otherwise, the different measures provide a coherent characterization of each wave mode motivating our choice to consider in the following channel-wise measures for the analyses, with the exception of the wave-wise planarity measure $P$, which has no channel-wise equivalent.

\subsection{Dataset comparisons quantify the variability of slow wave characteristics}
\label{subsec:results_dataset_comparison}
Based on the Cobrawap implementation, we are now in a position to perform quantitative comparisons of slow wave dynamics across the described ECoG and wide-field calcium imaging datasets, contrasting various experimental parameters. In the following, we demonstrate the application of the pipeline to investigate the influences of the anesthetic type and dosage, the application of disease models via genetic knock-out, and the measurement technique itself, in particular, its spatial resolution.

\begin{figure}
  \begin{center}
    \includegraphics[width=\textwidth]{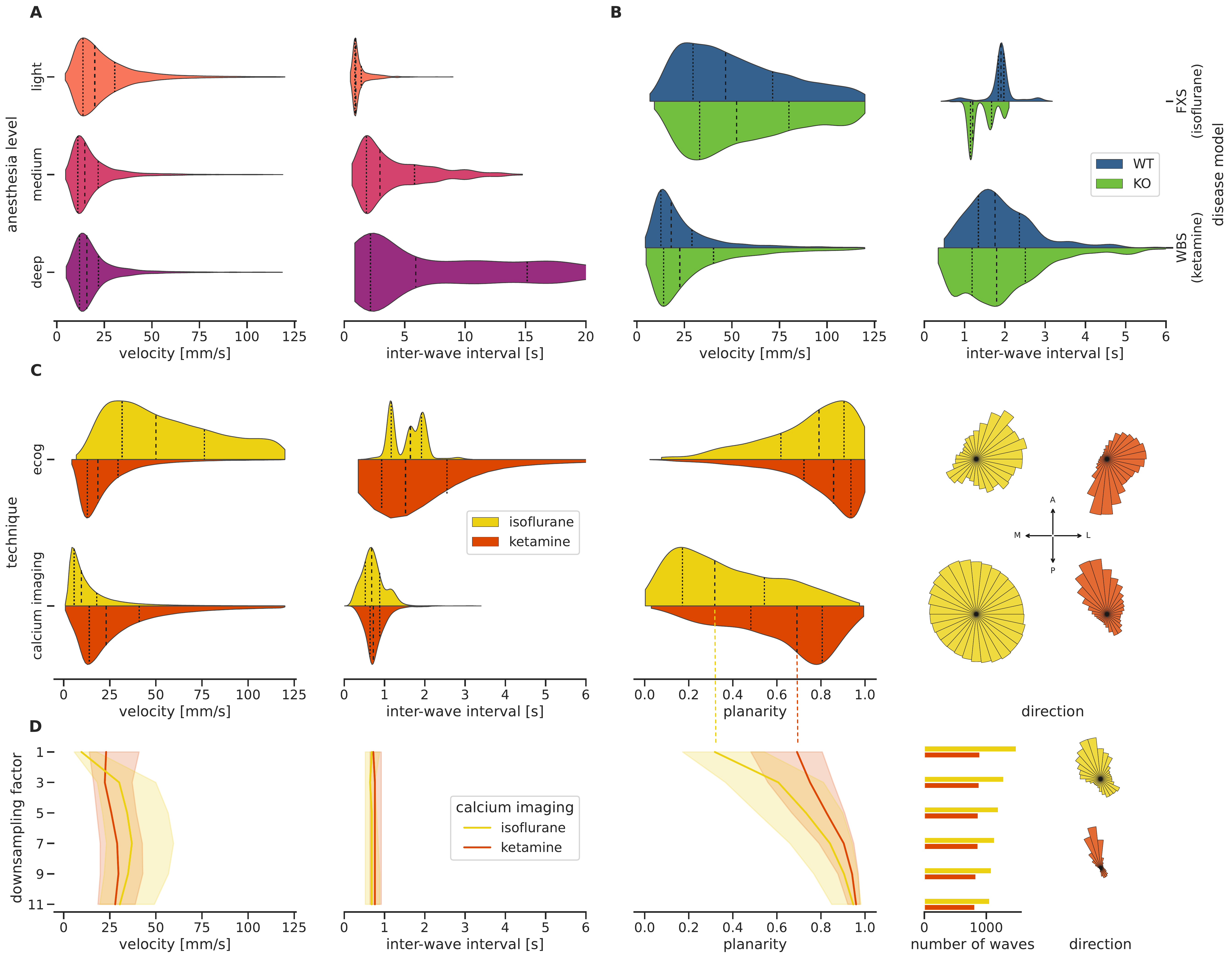}
  \end{center}
  \caption{\textbf{Quantitative comparison of slow waves across heterogeneous datasets.} Violin plots show sample distributions with indications of the median (dashed line) and the quartiles (dotted lines). Line plots also show the median and quartiles (shaded areas). Polar plots show the distributions of wave directions in the right hemisphere so that "up" corresponds to an anterior direction and "right" to a lateral direction. \textbf{A}: Velocity and inter-wave intervals of slow waves in ECoG recordings as a function of the anesthesia level. \textbf{B}: Velocity and inter-wave intervals of slow waves in ECoG recordings of experiments modeling Willems-Beuren Syndrome (WBS) and Fragile-X Syndrome (FXS) split into wild-type (WT, blue) and knock-out (KO, green) subjects. \textbf{C}: The ECoG data from panel B is compared to calcium imaging data, split into anesthetic types, on the basis of wave velocity, inter-wave interval, wave planarity, and wave direction. \textbf{D}: Effect of stepwise spatially downsampling the calcium imaging data from $0.05$~mm (factor 1) to $0.55$~mm (factor 11, the spatial resolution of the ECoG data) on wave velocity, inter-wave interval, wave planarity, number of waves, and wave direction. The histograms of wave directions are only shown for the fully downsampled data (with factor 11).}
  \label{fig:dataset_comparison}
\end{figure}

To check the validity of the pipeline, we first qualitatively replicate results that were previously published using the same datasets. It has been shown that the dynamics of slow waves crucially depend on the level of anesthesia. While the velocity of waves tends to decrease slightly in deeper anesthesia states \citep{Pazienti2022_103918}, the inter-wave intervals become more prolonged, i.e., the frequency of waves decreases \citep{Pazienti2022_103918, Dasilva2021_117415}. The same trends are visible in the corresponding pipeline output for the same data (\prettyref{fig:dataset_comparison}A). The velocity and frequency of slow waves were also measured in the context of a disease model for Williams-Beuren Syndrome (WBS) in knock-out (KO) condition and wild-type (WT) (of the same genetic strain) \citep{Dasilva2020_765, Sanchez-Vives2019_a}. In both, the previous publication and the pipeline output (\prettyref{fig:dataset_comparison}B), we observe no visible effect on the wave characteristics except for a slight increase in the variance in the knock-out condition.

Including another dataset from an experiment \citep{Sanchez-Vives2019_} that models the Fragile-X Syndrome (FXS) allows us to extend the analysis of the WBS data across experiments. Focusing on the wild-type control subjects, we compare the influence of experimental parameters between the WBS and FXS experiments \prettyref{fig:dataset_comparison}B). A notable difference between the two experimental setups is that in the WBS experiment, ketamine was used as the anesthetic ($100$~mg/kg inducing + $37$~mg/kg maintaining), while the FXS experiment used isoflurane ($4$~\% inducing + $1$~\% maintaining). Comparing the distributions of wave propagation velocities for the wild-type mice shows considerably larger velocities measured in the experiment that used isoflurane and a larger range of inter-wave intervals for the experiment that used ketamine (\prettyref{fig:dataset_comparison}B). In comparison to \prettyref{fig:dataset_comparison}A, where anesthesia was induced with ketamine ($75$~mg/kg) but maintained with isoflurane ($0.1$-$1.16$~\%), we see a better agreement to the velocities in the WBS (ketamine) experiment than to the FXS (isoflurane) experiment. However, comparing the exact depth of anesthesia across different anaesthetics is generally difficult. Furthermore, it is to be noted that since this is a meta-analysis, there is little control for confounding parameters between the different datasets. So, care must be taken in the attribution of the differences in wave characteristics to a single parameter, here the anesthetic type.

Next, we broadened the scope of the analysis by contrasting the ECoG recordings of ketamine- and isoflurane-anesthetized mice to analogous recordings that use wide-field calcium imaging on anesthetized Thy1-GCaMP6f mice to measure the cortical activity via the fluorescent response in excitatory neurons \citep{Resta2020_, Resta2020_a}. 
\prettyref{fig:dataset_comparison}C illustrates the distributions of wave characteristics grouped by measurement technique and anesthetic type. A principal difference between the measurement techniques is their spatial resolution. The wide-field calcium imaging data has a resolution of $0.05$~mm compared to $0.55$~mm for the ECoG data. The finer resolution allows for a better distinction of complex non-planar wave patterns, as can be seen by the broader distribution of the planarity that is shifted towards smaller values. Additionally, in calcium imaging data complex wave patterns with low planarity are more prevalent under isoflurane-induced anesthesia than under ketamine-induced anesthesia, an effect that can also be seen to a smaller extent in the ECoG recordings. Furthermore, the detected waves in the calcium imaging data are more frequent and regular, as shown by the inter-wave-interval distributions. The wave velocity distributions exhibit a notable discrepancy between the measurement techniques for the isoflurane datasets, while the velocities for the ketamine dataset are quite similar. This considerable difference in wave velocities is likely related to a difference in the isoflurane concentration ($1\%$ in ECoG and $1.5-2\%$ in calcium imaging recordings), as even small differences in the concentration can have a considerable effect on the wave dynamics (cf.~\prettyref{fig:dataset_comparison}A,B).

The slow waves we detect with the Cobrawap pipeline tend to propagate along a preferred axis and primarily in one direction. This axis seems to be approximately consistent within the data of each measurement technique but not across (\prettyref{fig:dataset_comparison}C, right). In the ECoG data, the preferred propagation axis spans from posterior-medial to anterior-lateral, with the preferred direction being different for the isoflurane and ketamine datasets. In the wide-field calcium imaging data, the preferred wave direction is from posterior-lateral to anterior-medial. Wave propagation that is oriented in a back-to-front or front-to-back manner is also reported in previous studies \citep{Sanchez-Vives2000_1027, Massimini2004_1160, Nir2011_153, Ruiz-Mejias2011_2910, Sheroziya2014_8875, Greenberg2018_328, Pazienti2022_103918}.
The spread of the wave direction histogram around the preferred directions can be either caused by a variance of channel-wise directions between waves or within waves, e.g., waves with low planarity have, per definition, a broader spread of channel-wise directions. 

To further explain the observed differences in the wave characteristics between the ECoG and calcium imaging data, we investigate the influence of their different spatial resolution by spatially downsampling the calcium imaging data up to a factor of $11$, for which the spatial resolution is equal to the one of ECoG ($0.55$~mm)
\prettyref{fig:dataset_comparison}D shows how the distributions of wave characteristics change as a function of the downsampling factor. With a decreasing spatial resolution, fewer waves are detected, and they appear more planar as some complex local patterns are no longer detected. This effect is particularly visible for the isoflurane datasets. A similar effect on the probability of detecting a planar wave as a function of ROI size has been previously shown by \citet{Liang2021_3665}. The histograms of directions of the fully downsampled calcium imaging data are more narrow than for the full resolution (\prettyref{fig:dataset_comparison}C). This indicates that the propagation directions are consistent across waves, and the variances in direction observed in \prettyref{fig:dataset_comparison}C are caused mainly by non-planar waves. Lastly, we observe that with increasing planarity as a result of increased downsampling, the waves in the isoflurane datasets exhibit faster channel-wise velocities that surpass the ketamine wave velocities, comparable to the ECoG data.

In summary, we demonstrate how the adaptable pipeline approach of Cobrawap enables the comparison of slow wave characteristics across heterogenous datasets, including electrical and optical acquisition methods. This meta-analysis illustrates distinct differences within the aggregated data and potential dependencies on the experimental parameters to be further investigated.

\subsection{Interchangeable blocks enable benchmarking of methods}
\label{subsec:results_method_comparison}
\begin{figure}
  \begin{center}
    \includegraphics[width=\textwidth]{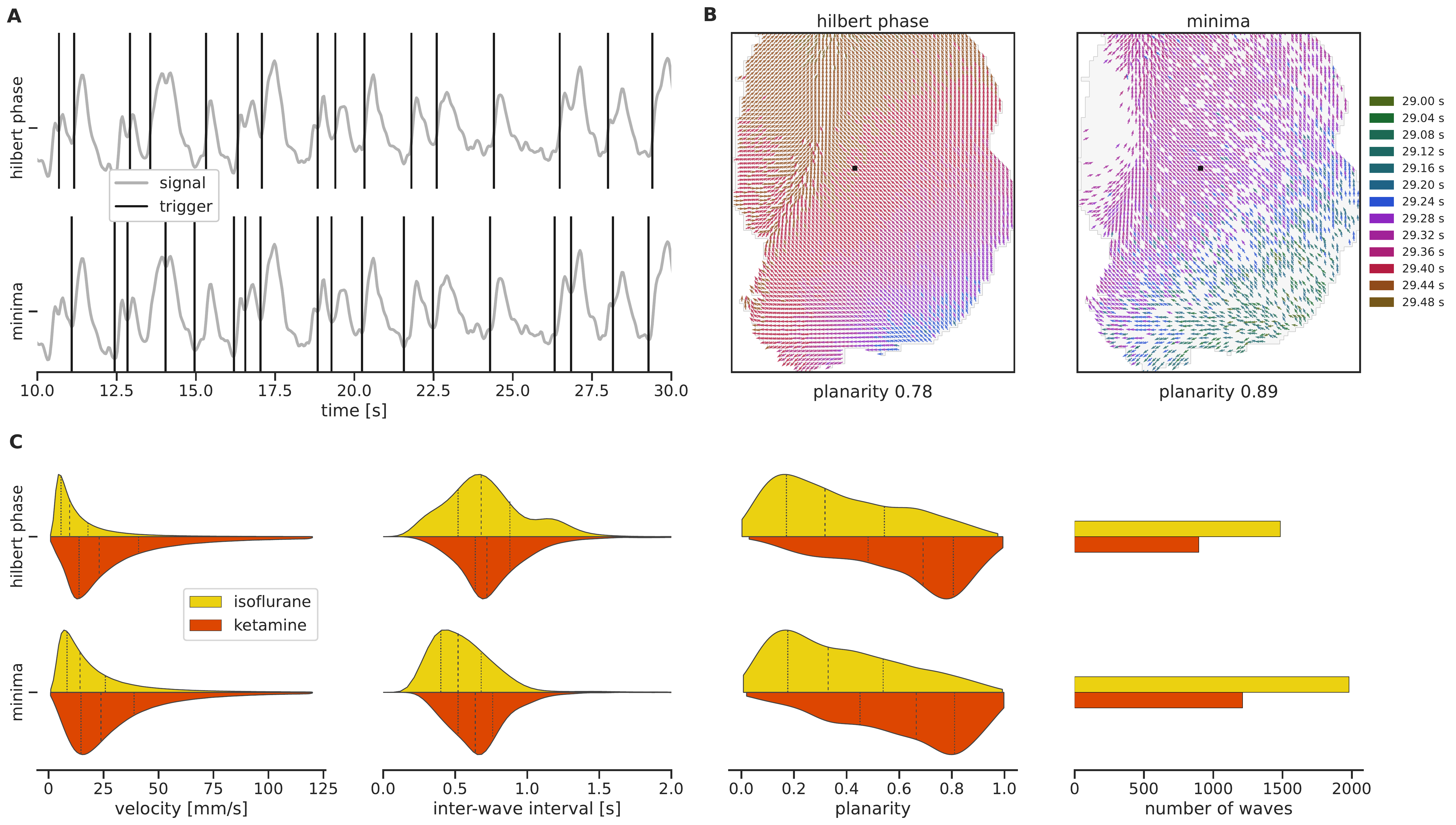}
  \end{center}
  \caption{\textbf{Comparison of trigger detection methods in calcium imaging data.} \textbf{A}: Upward transitions (triggers, black vertical lines) found by two different detection algorithms, detecting crossing of the Hilbert phase at $- \pi / 2$ (top) and local minima (bottom). Signal taken from the black pixel indicated in panel B. \textbf{B}: The same exemplary wave (corresponding to the last trigger in panel A) illustrated over the recorded area as detected using the two trigger detection algorithms. The arrows indicate the local direction of the wavefront at the time of the trigger, which is also encoded as the color of the arrows. \textbf{C}: The distributions of the wave velocity, inter-wave interval, wave planarity, and number of the waves obtained by the two methods in the calcium imaging datasets (compare to \prettyref{fig:dataset_comparison}C).}
  \label{fig:method_comparison}
\end{figure}

While applying the same analysis method to different data enables rigorous comparisons, applying alternative methods to the same data allows investigating the influence of the choice of the method itself. In the analysis of slow waves, the method for detecting the transitions from Down to Up states plays  a central role that we will consider as an example in the following. So far, in the calcium imaging data we detected the trigger times at the upstroke of the transitions as the Hilbert phase of the signal crossing a threshold value of $-\frac{\pi}{2}$ (see \prettyref{subsec:methods_trigger_detection}). However, alternative methods to define trigger times were suggested, such as using the local minima of the filtered signal \citep[cf., e.g.,][]{Celotto2020_14}. \prettyref{fig:method_comparison} illustrates the influence of the two different detection methods on the resulting wave characteristics. In Cobrawap, realizing this workflow for benchmarking the two methods with our pipeline only requires selecting the corresponding wave detection block and to rerun the analysis on the calcium imaging data. As shown in The detected triggers differ clearly in number and exact timing (\prettyref{fig:method_comparison}A), resulting in a different set of detected waves (see an example in \prettyref{fig:method_comparison}B). \prettyref{fig:method_comparison}C shows that the total number of waves is larger with the minima method ensuing that the corresponding inter-wave-intervals also tend to be short then for the Hilbert-phase method (effect size $0.43$ for ketamine, $0.58$ for isoflurane). The velocities remain similar for the ketamine datasets (effect size $0.04$) but differ slightly for the isoflurane datasets (effect size $0.32$) while he planarity distributions are mostly unaffected by the choice of trigger detection method (effect size $0.03$ for ketamine and $0.02$ for isoflurane). A Kolmogorov-Smirnov test indicates significant ($p<0.01$) differences for the velocity and inter-wave-interval, but not the planarity.

The ability to easily compare methods allows us to evaluate the strengths of each approach of detecting upward transitions and check for potential biases introduced to the wave characterization. The minima detection method is less strict and therefore detects more waves, including some smaller local ones. The Hilbert-phase method detects fewer waves, which are, however, better separated and more coherent across channels. For a more extensive method comparison, including specific edge cases, this approach could be further combined with simulated data.

\section{Discussion}
\label{sec:discussion}

\subsection{Advantages of a reusable modular pipeline  design}
The presented multi-modal analyses of slow wave activity using the Cobrawap implementation illustrate the benefits of a modular pipeline approach that incorporates general aspects of reproducibility and reusability.
The pipeline output retains information about the applied analysis scripts, their execution order, and parameter settings. The intermediate stage and block results and their visualizations further help to retrace the workflow and build confidence in the findings.
Aligning the workflows for different datasets by applying the same or analogous analysis methods while catering to their specific processing demands makes the corresponding results comparable. This setup promotes cross-domain comparisons, including the quantitative evaluation of experimental parameters (e.g., measurement techniques, anesthetics, species) and validation of simulated activity data.
The modular nature of the pipeline design can cater to heterogeneous data inputs. Additionally, interchanging methods in the analysis of the same dataset also allows the evaluation of a method's influence on the downstream results.
The explicit extensibility of the pipeline and reusability of the individual components aims to facilitate further research applications by providing a framework for designing efficient, and reproducible workflows.

\subsection{Structured analysis pipelines contribute to progressing the study of slow waves}
The presented meta-analysis across heterogenous datasets comprises ECoG and wide-field calcium imaging recordings (\prettyref{fig:dataset_comparison}). These measurement techniques are known to result in fundamentally different signals. While ECoG tends to record only the spiking activity of neurons in the superficial layer with a high firing rate and a high signal-to-noise ratio with high temporal and low spatial resolution, wide-field imaging of GCaMP6f in Thy1-GCaMP6f mice measures population spiking activity from excitatory neurons in layers 2/3 and 5 as a delayed, low-pass filtered, non-linearly transformed fluorescence signal with low temporal and high spatial resolution \citep{Siegle2021_e69068, Siegle2021_86, deVries2020_138}. Therefore, even elaborate models can not fully capture all aspects of the complex relationship between these two measurement techniques, and there is generally no precise agreement between results  beyond coarse qualitative measures \citep{Chen2013_295b, Wei2020_e1008198, Stringer2019_361}. In this context, Cobrawap can quantitatively illustrate the differences in the measurement types regarding the characteristics of slow wave activity. For analyzing wave characteristics, it is of particular interest from which cortical layer the measurement technique samples the contributing neurons since aspects like frequency power or propagation speed are known to vary considerably with cortical depth \citep{Halgren2018_2055, Capone2019_319}. 
Taking these considerations into account, an integrative approach of using multiple measurement techniques may benefit from the complementing viewpoints that, for example, ECoG and calcium imaging can provide.

Besides the biases of the measurement technique and its resolution, we further present the influences of the anesthetic type and dosage on the wave characteristics, showing in particular that ketamine tends to produce more planar waves than isoflurane, in turn also influencing the measured directions and velocities. This effect is likely linked to the known attributes of the anesthetics, that ketamine is more effective in generating slow-wave activity as it increases LFP power in the delta frequency band while isoflurane rather enhances LFP activity in the theta band and above \citep{Michelson2018_2232, Purdon2015_937}.

The need to quantitatively relate results from the literature to each other becomes quite apparent when, for example, investigating the sources of variance of the velocity of slow waves, which can vary from a few mm/s in recordings of anesthetized rodents up to $\sim10$~m/s in human sleep experiments \citep{Massimini2004_1160, Ruiz-Mejias2011_2910, Muller2016_e17267}. Studied influences to this variability include the extent of axonal projections \citep{Golomb1996_750, Compte2003_2707, Massimini2004_1160}, axonal conductances \citep{Ruiz-Mejias2011_2910}, involved cell types \citep{Bazhenov2002_8691}, and neuronal excitability depending on anesthetics \citep{Pazienti2022_103918}, neuromodulators \citep{Destexhe1999_427}, or cortico-cortical or cortico-thalamic loops \citep{Sanchez-Vives2000_1027, Mohajerani2013_1426a}. 
Furthermore, the velocity of a wave may depend on its direction, which in turn is influenced by an interplay of local and global connectivity properties and frequency effects \citep{Massimini2004_1160, Mohajerani2013_1426a, Galinsky2020_023061}. Comparison between data from different studies can help relate and discern such influences. 

\subsection{Integration in model development and data-driven simulations}
While the exploration of wave characteristics under different conditions can provide further insight into the understanding of the underlying processes, availability of experimental data can also suffer from constraints in the data size, parameter regime, and uncontrolled confounds. Therefore, in many scenarios, it is beneficial to include modeling data in the analysis.
Insights from experimental analyses can inform the model development process to adequately describe the biological reality.
However, Cobrawap can also be directly applied to simulation outcomes to extract the same characteristics as from the experimental data and perform a quantitative comparison in a subsequent validation step. 
Such a step can already be integrated into the model development in the form of an explicit calibration. This strategy is considered in \citet{Capone2022_a} where, after a preliminary estimation of model parameters through likelihood maximization \citep{Capone2018_17056}, a subset of parameters is further adjusted by performing a grid exploration relying on the direct comparison between data and simulations based on Cobrawap. The comparisons in \cite{Capone2022_a} are derived from the selection of observables described here, considering the waves’ local velocities, directions, and frequencies. This calibration approach allowed for a meta-inference procedure, finding the optimal parameters of a neuromodulation current to reproduce the dynamics observed in experimental data. Such an approach is essential to complement the theoretical understanding of the relationship between the spatio-temporal features of cortical waves and the cortical structure \citep{Capone2017_39611}.

\subsection{Reusability: related pipelines and outlook}
We developed the Cobrawap pipeline to be reusable. Its modular structure of stages and blocks allows for reuse in different scenarios. The pipeline may be applied to other types of input data, may be extended by other method blocks, or changed to produce additional kinds of output.
The pipeline can be adapted in this regard by editing the stage's config files and changing the block selection and parameter settings. The minimum requirement for any input data is that it is recorded on a grid electrode/pixel layout.
Cobrawap can be extended for more substantial changes by adding new blocks that implement specific analysis methods. Further, disparate applications may swap out the later stages of the pipeline entirely, i.e., realizing a branching-off pipeline (similar to the separate stage 5 realizations for channel-wise and wave-wise observables).
The individual blocks and stages can also be used selectively as stand-alone elements without the pipeline in different workflow applications.

For example, current work entails a more detailed analysis of the local oscillations ignoring the spatial propagation, similar to work done in \citep{DeBonis2019_70}. This application reuses the first three stages of Cobrawap and then branches off with specialized stages. 
Another envisioned pipeline derivation would focus on data that shows oscillations as a result of electrical stimulation. This application would need additional specialized blocks, for example, to quantify response complexity by a perturbation-complexity index (PCI).
There are also wave-like phenomena in other frequency regimes. For example, in alpha, beta, and gamma frequency ranges, diverse wave patterns have been observed in awake behaving animals \citep[e.g.][]{Senseman2002_1499, Rubino2006_1549, Townsend2018_e1006643, Denker2018_1, Davis2020_1}. A flexible pipeline approach following Cobrawap may disentangle some of the reported results, methods, and terminologies.

\subsection{Conclusion}
In this paper, we demonstrate the advantages of formalizing and harmonizing analysis approaches. By taking a data-science perspective, we work towards integrating heterogeneous insights from different data and analysis types. In our view, understanding an organ as complex as the brain requires the integration of data obtained on multiple levels of observation. Furthermore, we experienced how structuring our methodology and implementation also contributed greatly to our structure of thought. We are confident that the concepts presented in the framework of the Cobrawap implementation contribute to advocating the concept of reusability for analysis resources, in particular with regard to the uptake of and contribution to community software projects.

\section{Materials and methods}
\label{sec:methods}

\subsection{Resource Availability}
\subsubsection{Lead Contact}
Further information and requests for resources should be directed to and will be fulfilled by the lead contact, Robin Gutzen (r.gutzen@fz-juelich.de, OrcID: 0000-0001-7373-5962).

\subsubsection{Materials Availability}
This study did not generate new unique reagents.

\subsubsection{Data and Code Availability}
\begin{itemize}
    \item This paper analyzes existing data that is publicly available via the EBRAINS Knowledge Graph. The DOIs are listed in the key resources table. 
    \item All original code has been deposited at Zenodo and is publicly available as of the date of publication. The DOIs are listed in the key resources table.
    \item Any additional information required to reanalyze the data reported in this paper is available from the lead contact upon request.
\end{itemize}

\subsection{Experimental model and subject details}
\label{subsec:methods_datasets}
An overview of all the individual recordings is presented in Appendix \prettyref{fig:dataset_overview}.

\subsubsection{Mouse ECoG recordings}
\label{subsubsec:datasets_IDIBAPS}
The three experimental ECoG datasets have been provided by IDIBAPS (Institut d’Investigacions Biomèdiques Agustí Pi i Sunyer): Williams Beuren Syndrome (WBS) 3-4 months old adult male mice (Wild-Type and Knock-Out), Fragile X Syndrome (FXS) (Wild-Type and Knock-Out) mice  and Propagation Modes of Cortical Slow Waves across anesthesia levels in adult male C57BL/6J mice (PMSW). All animals were bred in-house at the University of Barcelona and kept under a $12$~h light/dark cycle with food and water \textit{ad libitum}. All procedures were approved by the Ethics Committee at the Hospital Clínic of Barcelona and were carried out to the standards laid down in Spanish regulatory laws (BOE-A-2013-6271) and European Communities Directive (2010/63/EU).

For WBS subjects, an intraperitoneal injection of ketamine ($100$~mg/kg) and medetomidine ($1.3$~mg/kg) was administered to induce anesthesia. It was maintained by a constant administration of subcutaneous ketamine ($37$~mg/kg/h). For FXS subjects, anesthesia was induced by the inhalation of $4$\% isofluorane in $100$\% oxygen for induction and $1$\% for maintenance. Finally, for PMSW subjects, an intraperitoneal injection of ketamine ($75$~mg/kg) and medetomidine ($1.3$~mg/kg) and maintained by the inhalation of different concentrations of isoflurane in pure oxygen. In PMSW, three levels of anesthesia were reached that were classified according to the provided isoflurane concentrations: deep=$1.16\pm0.08\%$ (s.e.m); medium=$0.34\pm0.06\%$; light=$0.1\pm0.0\%$. The volume delivered was $0.8$~L/min. 

In order to avoid respiratory secretions and edema, atropine ($0.3$~mg/kg), methylprednisolone ($30$~mg/kg), and mannitol ($0.5$~g/kg) were administered subcutaneously to all subjects. So as to aid breathing and once in the surgical plane of anesthesia, a tracheotomy was performed. The animal was then placed on a stereotaxic frame (SR-6M, Narishige, Japan) with constant body temperature monitoring maintained at $37^\circ$C by means of a thermal blanket (RWD Life Science, China). A wide craniotomy and durotomy were performed over the left or right (only left in FXS) hemisphere from -$3.0$~mm to +$3.0$~mm relative to the bregma and +$3.0$~mm relative to the midline. A 32-channel multielectrode array ($550~\mu$m spacing) covering a large part of the hemisphere’s surface was used to record the extracellular micro-electrocorticogram (micro-ECoG) activity. 
For WBS and FXS datasets, recordings were acquired from spontaneous activity in the animal under anesthesia. Regarding the PMSW  dataset, each anesthesia level was maintained for $20$–$30$ minutes, and spontaneous recordings were consistently obtained in a stable slow oscillatory regime (approximately $10$ minutes after the change in concentration). During the recording protocol, a precise visual inspection of all channels was made in order to ensure that all of them were properly acquiring the signal.

The signals were amplified (Multichannel Systems, GmbH), digitized at $5$~kHz, and fed into a computer via a digitizer interface (CED 1401 and Spike2 software, Cambridge Electronic Design, UK).

\subsubsection{Mouse wide-field calcium imaging recordings} 
\label{subsubsec:datasets_LENS}
Experimental data acquired from mice have been provided by LENS (European Laboratory for Non-Linear Spectroscopy\footnote{LENS Home Page, \url{http://www.lens.unifi.it} (accessed on Nov. 2019)}) and by the Department of Physics and Astronomy of the University of Florence. All procedures involving mice were performed in accordance with the rules of the Italian Minister of Health (Protocol Number 183/2016-PR). 
Mice were housed in clear plastic enriched cages under a $12$~h light/dark cycle and were given \textit{ad libitum} access to water and food. 

Mouse Model: The transgenic mouse line used is the C57BL/6J-Tg(Thy1GCaMP6f)GP5.17Dkim/J (referred to as GCaMP6f mice\footnote{For more details, see The Jackson Laboratory, Thy1-GCaMP6f, \url{https://www.jax.org/strain/025393} (accessed on Nov. 2019).}) from Jackson Laboratories (Bar Harbor, Maine USA). In this mouse model, the ultra-sensitive calcium indicator (GCaMP6f) is selectively expressed in excitatory neurons \citep{Chen2013_295, Dana2014_e108697}.

Surgery and wide-field imaging: Surgery procedures and imaging protocols were performed as described in \citep{Celotto2020_14}. Briefly, 6 months old male mice are anesthetized with either a mix of ketamine and Xylazine in doses of $100$~mg/kg and $10$~mg/kg respectively or isoflurane ($3-4\%$ induction and $1.5-2\%$ maintaining). 
To obtain optical access to neuronal activity over the right hemisphere, the local anesthetic lidocaine ($20$~mg/mL) was applied and the skin and the periosteum over the skull were removed. Wide-field imaging was performed right after the surgical procedure.  
GCaMP6f fluorescence imaging was performed with a $505$~nm LED light (M505L3 Thorlabs, New Jersey, United States) deflected by a dichroic filter (DC FF 495-DI02 Semrock, Rochester, New York, USA) on the objective (2.5x EC Plan Neofluar, NA 0.085, Carl Zeiss Microscopy, Oberkochen, Germany). The fluorescence signal was selected by a band-pass filter (525/50 Semrock, Rochester, New York, USA) and collected on the sensor of a high-speed complementary metal-oxide semiconductor (CMOS) camera (Orca Flash 4.0 Hamamatsu Photonics, NJ, USA). 

\subsection{Method details}
\subsubsection{Design of the analysis pipeline}
\label{subsec:methods_pipeline}
\paragraph*{Code development}
The implementation of the "Collaborative Brain Wave Analysis pipeline" (Cobrawap) infrastructure is being developed on GitHub\footnote{\url{https://github.com/INM-6/cobrawap}}.
The pipeline configuration for the presented pipeline application and additional analysis and plotting code is stored in a separate GitHub repository\footnote{\url{https://github.com/INM-6/slow_wave_analysis}}.

The implementation of the general "Collaborative Brain Wave Analysis pipeline" (Cobrawap) infrastructure is being developed on GitHub\footnote{\url{https://github.com/INM-6/cobrawap}}.

\paragraph*{Terminology}
We organize the analysis pipeline hierarchically into three layers. The top layer constitutes the pipeline itself or a task-specific realization of a pipeline, which we here call \textit{workflow}. A \textit{pipeline} we define as a sequence of processing/analysis stages to be executed following a given order (“from left to right”). As a \textit{stage} we describe a self-consistent logical episode within the analysis process, such that the output of a stage can be considered a reasonable intermediate result. Furthermore, a stage should be general enough to be reusable in multiple workflows or pipelines. Each stage is segmented into blocks, which can be selected and rearranged depending on the configurations of the user and the mechanics of the stage. A \textit{block} is the smallest unit of the analysis pipeline and performs a specific action on the data. Blocks implement methods. In the case of alternative methods or alternative algorithms implementing a method, they can be either represented as options of a single block or separate blocks. The related terms “action”, “step”, “method”, and “algorithm”, we use without special definition in their common sense.

\paragraph*{Implementation with snakemake}
We designed the structure of the pipelines having in mind the features of the Snakemake workflow management framework \citep{Molder2021_33}. The rules are defined in script files called \textit{snakefile} which also link to a config file. Thus, our pipeline structure is conveniently mappable onto the snakemake elements: blocks are represented by rules and stages by snakefiles. In addition, we use another top-level snakefile to combine the stages as snakemake \textit{subworkflows} and make the pipeline executable as a whole. Within the stage Snakefiles, each block is represented by a snakemake rule which in most cases executes a Python script.
Furthermore, we expand the standard functionality of snakemake by three mechanics required by our pipeline design: 1) chaining the stages by linking the outputs and inputs of subworkflows, 2) manually selecting a specific block (i.e., method) or a sequence of blocks by choosing the desired methods in a config file, and 3) selecting and switching between sets of configs files (\textit{"profiles"}) for all stages. More details of the pipeline implementation are further explained in \prettyref{subsec:methods_pipeline}.

\paragraph*{Modularity}
One of the main design principles in constructing the analysis pipeline is modularity. This has the purpose of making the pipeline flexible and thus adaptable to different demands, by making it possible to rearrange and switch elements of the pipeline. In contrast to other typical analysis workflows, here, the construction of a specific workflow does not require the changing of any scripts but is rather like tracing a path along the selected stages and blocks within a larger framework offered by the pipeline. Practically, for the stages, this means that different combinations or variations of stages can be chained together. For the selection blocks, there are two flavors of modularity used in the stages: \textit{choose one}, selecting one method block from multiple options; and \textit{choose any}, selecting any number of method blocks in any order (see \prettyref{fig:pipeline_illustration}).
Another aspect of modularity is that each element should be usable on its own as well as in combination with other elements. Therefore, much care needs to be put into managing the respective interfaces where the elements interact.

\paragraph*{Pipeline stages}
For the analysis of slow wave activity, we chose five stages (\prettyref{fig:pipeline_illustration}) starting from more generic stages (Data Entry, Processing) to task-specific stages (Trigger Detection, Wave Detection, Wave Characterization) which build up the Collaborative Brain Wave Analysis Pipeline (Cobrawap):
\begin{enumerate}
    \item[1] \textbf{Data Entry:} 
    This first stage loads a dataset and the required and optional metadata and converts the data into a standardized representation scheme (using the Neo data format). This loading script is the only custom code that is required to add a new data source to the pipeline, integrating information from a data file and a corresponding config file. It is checked whether the resulting data object conforms with the requirements of the pipeline and an overview of a data sample is plotted.
    \item[2] \textbf{Processing:}
    In the second stage, the data is prepared for analysis. The user can select any combination of processing blocks to fit the data type and their analysis objectives. Where available, the blocks use standard function implementations by the Elephant Electrophysiology Analysis Toolkit \citep{Denker2018_a}, the stack of scientific Python packages (i.e. scipy, scikit, etc.), or algorithms from the literature. 
    \item[3] \textbf{Trigger Detection:}
    Based on the processed data, this stage detects the transition times from Down to Up states (upward transitions, i.e., trigger) and, if possible, Up to Down states (downward transitions) by applying one of the available trigger detection blocks. What this trigger exactly relates to depends on the dataset, the processing, and the detection method. Additionally, there are optional filter blocks that can be applied to clean the collection of detected triggers. The trigger collection is added as a \texttt{neo.Event} named 'transitions' to the input Neo object containing the processed data.
    This stage is general enough to also be of use for the analysis of other wave-like activity, beyond slow waves.
    \item[4] \textbf{Wave Detection:}
    Latest at this stage, the wave description converges to a common level. The selected detection method operates on the trigger times, grouping them into individual wavefronts while being completely agnostic about the type and origin of the original data. The resulting groups of triggers, i.e. waves, are added as another \texttt{neo.Event} named 'wavefronts'. Optionally, any number of additional wave descriptions can be calculated and added to the neo object, including the optical flow vector field or a wave-mode clustering.
    \item[5a] \textbf{Wave-wise Characterization:}
    The final stage calculates one or multiple characteristic measure(s) of the detected waves. This contains scalar measures as, for example, the wave velocity or its duration, but may also contain metadata information like analysis parameters or information about the dataset added in stage 1 (selected via the \texttt{annotations} block). The output is a pandas dataframe \citep{McKinney2010_56} where each row represents one wave and each column an attribute/characteristic. This pipeline output for one dataset can be directly merged or compared with the output for other datasets and serves as the basis for various cross-domain comparisons (e.g., data comparisons, model validation, method benchmarking).
    \item[5b] \textbf{Channel-wise Characterization:}
    This alternative final stage is equivalent to the 'Wave-wise Characterization' in its functionality, but its characteristic measures are calculated per wave and channel (i.e. electrode or pixel). Therefore, in the output dataframe, one row represents one channel for one wave.
    For either of the two options for the final stage, the characterization can also optionally be performed only on the wave modes instead of on each wave.
\end{enumerate}

\begin{figure}
\begin{fullwidth}
    \begin{center}
        \includegraphics[width=1.25\textwidth]{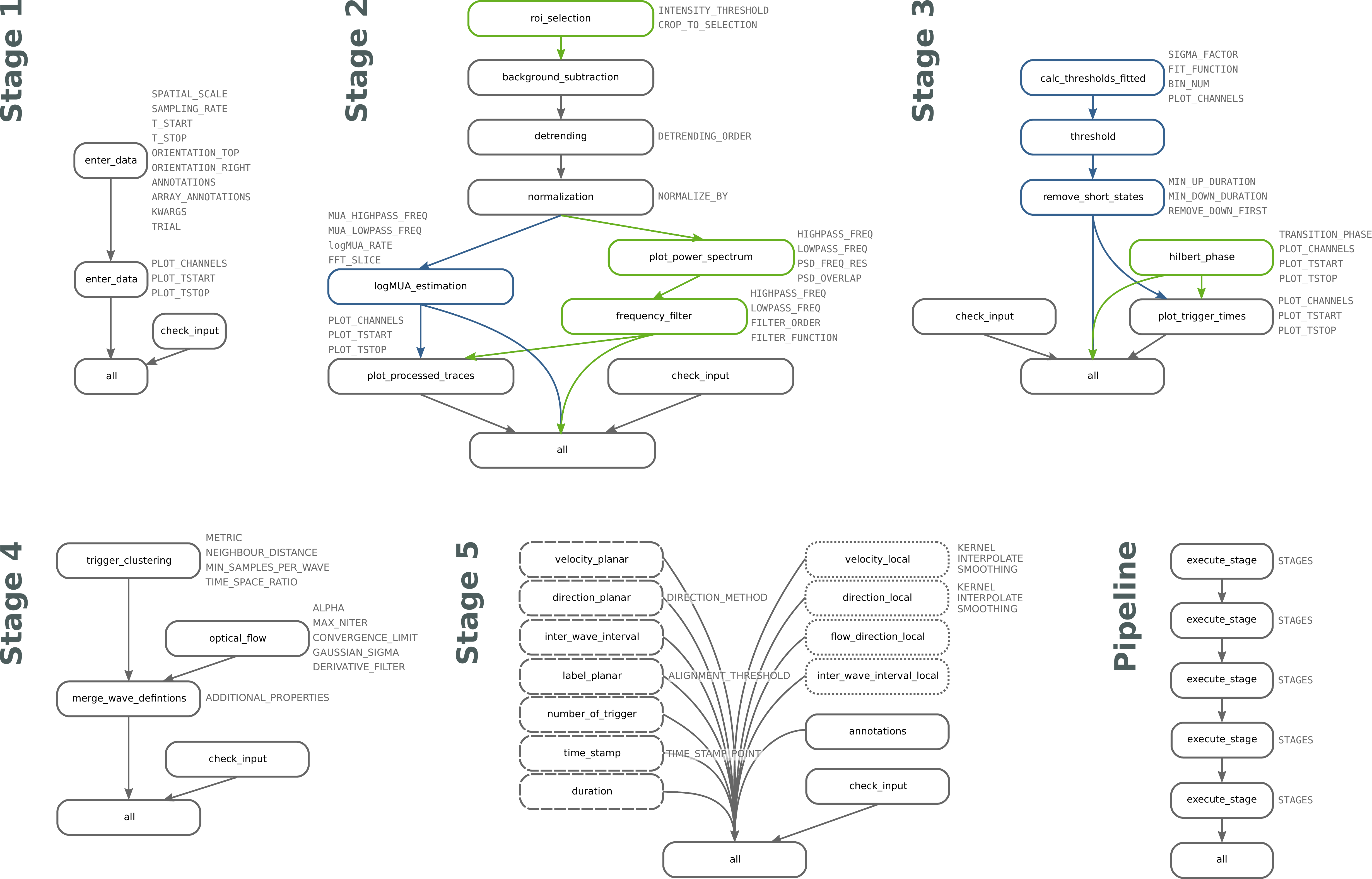}
    \end{center}
    \caption{\textbf{Pipeline structure.} The diagrams show the execution order of the blocks for each stage (plus the full pipeline) as generated by the snakemake workflow management framework \citep{Molder2021_33}. Blocks that are specific for the ECoG data are shown in blue, and blocks specific for calcium imaging in green, while common blocks are grey. In stage 5, dashed blocks are wave-wise measures and dotted blocks are channel-wise measures. Next to the blocks, the parameters are indicated that can be set in the corresponding config files.}
    \label{fig:snakemakeogramm}
\end{fullwidth}
\end{figure}

\paragraph*{Data and metadata representation}
When designing a pipeline with the objective of modularity and generality, it is of crucial importance to properly define the interfaces between the individual analysis elements (blocks, stages) as well as to the user and other tools.
This entails the representation of the data and metadata in a standardized format. For this, we chose the data format Neo \citep{Garcia2014_10}. Neo supports a variety of data types and reading and writing of various common file formats. This interoperability is, thus, ideal for aiding the flexible use of the pipeline.
Since Neo itself is very versatile, there are multiple ways how to organize the data and metadata in the Neo structure, so we need to be even more precise in standardizing the data structure. That means that within the pipeline we store the data of all channels in one `neo.AnalogSignal` object and the metadata in the corresponding annotations and array annotations for channel-wise metadata (like their x and y coordinates). Processing and transformation blocks overwrite the data in this Analogsignal object and add corresponding metadata. In stages 3 and 4, additional `neo.Event` objects may be added to represent transition times and wavefronts as well as an additional AnalogSignal object for derived vector fields (e.g., the optical flow).
The file format to use for storing the intermediate results of blocks and stages can be format supported by Neo. We recommend Nix \citep{Stoewer2014_} for a robust file format, or the pickle or numpy for a less robust format that is, however, faster to read and write and produces smaller files.

The entire first stage is dedicated to being the interface between the pipeline and the data resource. It checks whether the data has the required capabilities and then organizes data and metadata into the Neo structure. For the analysis of slow waves with this pipeline, the data needs to be obtained from electrodes or pixels that are arranged on a rectangular grid (which may include empty sites), and that exhibit propagating Up states. The corresponding minimal set of metadata required for the pipeline to process the data are i) the sampling rate, ii) the distance between the electrodes/pixels, iii) and their relative spatial locations of the grid as integer x and y coordinates.
Although not explicitly used, it is strongly recommended to include more information such as the measured cortical location, the spatial scale of the grid, the units of the signal, the type and dosage of the anesthetic, an identifier of the dataset, etc. This additional metadata is propagated through the pipeline alongside the data in order to reasonably use and interpret the results. 

\paragraph*{Pipeline interfaces}
This degree of flexibility in the execution order of both stages and blocks is based on standardizing the input and output formats. By defining the input requirements for each stage and block, they can successfully interact while remaining interchangeable and thus reusable for other pipelines or applications.
Since the individual stages are designed to be potentially reused in other pipelines, the stage outputs, i.e. the intermediate results, should suffice to the same level of completeness and documentation as a final result. Thus, also each stage needs to come with a detailed definition of its input and output structure which is checked by a dedicated 'check\_input' block. These definitions are collected in the stage's README file to guide developers of alternative pipelines as well as contributors of new blocks for the stage. 
Similarly, the individual blocks are also thought to serve the modular design by being easy to reuse and recombine, or even used as a standalone application. Therefore, they also need to clearly state the type and format of their in- and outputs. Other than for the stages, this is largely handled organically in form of the dependencies of the corresponding snakemake rule and the definition of the script's command line arguments and complemented by its docstring.

\paragraph*{Logging and intermediate results}
The modular organization of the pipeline facilitates maintainability, and additional built-in means, such as provenance tracking and storing intermediate results alongside their config files, further support reproducibility, and transparency. Moreover, we emphasize the integration of automatically generated plots of intermediate results. Most blocks produce a plot illustrating their function to make the evolution of the results (or potential bugs) visible.
To further enable the provenance of the analysis results, snakemake provides logs and reports which contain execution statistics, dependency trees, plots, and config settings. Additionally, we are currently working on integrating a formalized provenance tracking with \textit{fairgraph}\footnote{\url{https://pypi.org/project/fairgraph/}}.

\paragraph*{Pipeline configuration}
The flip side of flexibility and adaptability is complexity and ambiguity. The many combinatorial possibilities need to be controlled by a user interface separate from the actual analysis scripts, e.g., what stages and blocks should be executed, in which order, and with which parameters. Config files (e.g., in csv, yaml, json format) offer human-readable access and control to a user to adapt and execute different variations of the pipeline. 
Thus, we assign one config file to each stage.
Consequently, blocks need to be implemented having generality in mind with any specification handled by corresponding parameters settings, given as command line arguments, i.e. within the pipeline via the config file. Even though this approach is initially more time-consuming, it does pay off in both the quality of the method implementation and its (re-)usability. Furthermore, the availability and aggregation of parameters allow for easier and more transparent calibration of the pipeline across blocks and stages.
Additionally, there is a top-level config file for the entire pipeline that specifies the stages and their order and can define global parameters that may also overwrite stage parameters, e.g., for setting the file format or plotting parameters for all stages. Parameters in the config files are typically calibrated for a specific data type or experiment setup. To conveniently switch between calibration presets, the pipeline supports a hierarchical organization of config presets via \textit{profiles}. By executing the pipeline with \texttt{PROFILE=data1}, for each stage the corresponding config file \texttt{config\_data1.yaml} is used. For more versatility, profile names can use underscores to define subcategories and exceptions, e.g., \texttt{data1\_subject3}. In this case, each stage first looks if a corresponding config file of the same name exists, and if not removes the subcategory with the last underscore from the name, and repeats this lookup until it finds the named config file or defaults to \texttt{config.yaml}. Furthermore, profiles can have variations indicated in the name with a '|', e.g., \texttt{data1\_subject3|methodA}. This variation key is not removed when first looking up existing config files in the naming hierarchy, only when \texttt{config|methodA.yaml} doesn't exit it is removed and the lookup loop is repeated.

\subsubsection{LogMUA estimation \textit{(in stage 2)}}
\label{subsec:logmua_estimation}
The multi-unit activity (MUA) is an estimate of the local population firing rate, based on the relative spectral power in the high-frequency regime ($200$-$1500$~Hz) of the extracellular recordings. \citep{Mattia2010_812, Reig2009_1253, DeBonis2019_70} The corresponding algorithm first selects a moving window that samples the recording at a given rate. From these samples, the power spectral density (PSD) is calculated using the Welch algorithm. The MUA is defined as the average power in the defined frequency band divided by the average power of the full spectrum. Using the logarithm of the MUA helps to emphasize further the bi-modality of the distribution in the presence of slow oscillations.
In the selection of the parameters for the algorithm, it is crucial to choose a moving window size large enough so that the chosen frequencies can be accurately estimated ($\mathrm{window\; size} \le \frac{1}{\mathrm{highpass\; frequency}}$) and a corresponding MUA rate so that the full recording is sampled from ($\mathrm{MUA\; rate} < \frac{1}{\mathrm{window\; size}}$). 

\subsubsection{Trigger detection \textit{(in stage 3)}}
\label{subsec:methods_trigger_detection}
The pipeline implementation provides multiple options to detect trigger events, i.e., transitions from a low activity state to a high activity state (Up).

\begin{itemize}
    \item[\textit{threshold:}] 
The trigger events can either be defined by setting a threshold value for all the signals or by fitting a bimodal function to the amplitude distribution for each channel in order to set the threshold value. In the latter case, the fitting function is the sum of two Gaussians and the threshold value is set to the central minima. This option is applied to the ECoG datasets in this paper.
As an alternative to a double Gaussian fit, there is also the option to only fit the first peak corresponding to the low activity state by only looking at the data left of the peak and defining the threshold as $\mathrm{mean} + \mathrm{std}\cdot\mathtt{SIGMA\_FACTOR}$ with a user-defined \texttt{SIGMA\_FACTOR}.
Since the thresholding method detects also the corresponding downward transitions, this block is usually paired with an additional block that removes Up and Down states that are too short, given user-defined minimal Up and Down durations.

\item[\textit{Hilbert phase:}] 
Instead of detecting threshold crossings on the actual signal, the upstrokes of the upward transitions can be detected by thresholding the phase signal of the corresponding analytic signal. An adequate threshold value is a matter of definition, here, we apply $-\pi/2$, which corresponds well to the beginning of the upstroke in the actual signal.
To be more robust, the algorithm only selects time points where the threshold is crossed from smaller to larger values and where the crossing is followed by a peak (phase $=0$). This option is applied to the calcium imaging datasets in this paper unless otherwise indicated.

\item[\textit{minima:}]
As a third option, we adapted and improved the minima detection method presented in \citep{Celotto2020_14}. This method relies on the assumption that in an adequately filtered signal that the existence of a local minimum followed by a peak of a certain height indicates the start of a upward transition. This is particularly suitable for recording techniques characterized by a fast characteristic rise time (i.e. comparable with the theoretical minimum time interval between the passage of two waves on a single channel, e.g. optical data). We improved this method by including some further refinement on trigger candidates. Under the assumption that only one minima candidate can lie between two "good" local maxima candidates, we impose that 1) local maxima candidates need to have a signal intensity higher than a relative threshold value, determined in a moving window; 2) local maxima candidates need to be separated by a minimum distance (associated with the characteristic frequency of the investigate phenomenon); 3) a local minima candidate needs to be followed by a monotonically rising signal for a defined time interval (also associated to the characteristic frequency of the investigated phenomenon). If more than one candidate minimum is found between two local maxima candidates, the last one before the following "good" maxima is selected.
\end{itemize}

\subsubsection{Trigger clustering \textit{(in stage 4)}}
\label{subsec:trigger_clustering}
Wavefronts are defined as clusters of trigger times in the three-dimensional space of the electrode arrangement (x,y) and samples in time (t). To run a clustering algorithm in this space, the units of the time dimension need to be translated to the units of the spatial dimensions. The ideal transformation factor (\texttt{TIME\_SPACE\_RATIO}) depends on the expected dynamic of the phenomena. A wave that propagates linearly with $v_0$ is best recognized in the cluster when the time dimension is transformed by a factor $v_0 /(\mathrm{sampling\; rate} \times \mathrm{spatial\; scale})$. Thus, if we expect a propagation velocity roughly in the order of $\sim 10-20 \frac{mm}{s}$ then the transformation factor for the calcium imaging data with sampling rate $25$~Hz and spatial scale $50 \mu$m is $\sim 8-16 \frac{pixel}{frame}$. Here, we choose a \texttt{TIME\_SPACE\_RATIO} of $11$ for the calcium imaging data which scales according to the spatial resolution to a factor of $0.25$ for the logMUA ECoG data with a sampling rate of $100$~Hz.
The clustering is performed by a density-based algorithm (\texttt{scipy.cluster.DBSCAN}), illustrated in Supplemental \prettyref{fig:trigger_clustering}. The additional parameters for this algorithm are the minimum number of samples (\texttt{MIN\_SAMPLES\_PER\_WAVE}) and the typical distance between neighboring sample points (\texttt{NEIGHBOUR\_DISTANCE}) and were determined by calibrating test recordings from both calcium imaging and ECoG data and scaled consistently with the spatial resolution.

\subsubsection{Optical flow estimation \textit{(in stage 4)}}
\label{subsec:methods_optical_flow}
The optical flow is the pattern of apparent motion in a visual scene, which here corresponds to the recorded signal on the recording grid. To estimate the optical flow of the spatial propagation of activation, we apply the Horn-Schunck algorithm with a quadratic penalty function and a 3x3 Scharr derivative filter on the phase of the signal (the alternative application using the signal's amplitude, as well as different derivative filters can be selected via the configuration). Although other penalty functions, i.e., the Charbonnier function, are more accurate, we found that here the simple quadratic function is sufficient. This observation is in agreement with \citet{Townsend2018_e1006643} who report good results for the near quadratic edge case of the penalty function. Their study also guided our choice of the parameter $\alpha=1.5$, determining the weight of the smoothness constraint over the brightness constancy constraint. The resulting vector field is smoothed by a Gaussian kernel which reflects the dimensions of the expected wave activity with respect to the spatial and temporal scale of the data.

\subsection{Quantification and statistical analysis}

\subsubsection{Kernel estimation}
The kernel estimations for the plotted distributions in \prettyref{fig:dataset_comparison} and \prettyref{fig:method_comparison} use \code{scipy.gaussian\_kde} with the default Scott's rule \citep{Scott2015_} as bandwidth method, except for the distributions of inter-wave intervals which use $0.2$ times the standard deviation as the kernel size.

\subsubsection{Velocity filter}
Since the channel-wise velocity measure can produce unreasonably high values when there are near identical time delays between spatially distant triggers, we cap the presented distributions at $120$~mm/s.

\subsection{Key resources table}
\label{subsec:key_resource_table}

\begin{table}[h!]
    \begin{fullwidth}
    \label{tab:key_resource_table}
    \caption{Key resources table} 
\begin{tabular}{lllll}
& REAGENT or RESOURCE & SOURCE & IDENTIFIER & \\
\midrule
& Deposited data  &  &  &  \\ \hline
& Mouse ECoG (WBS) & EBRAINS Knowledge Graph & \doi{10.25493/DZWT-1T8} & \\
& Mouse ECoG (FXS) & EBRAINS Knowledge Graph & \doi{10.25493/ANF9-EG3} & \\
& Mouse ECoG (Propagation modes) & EBRAINS Knowledge Graph & \doi{10.25493/WKA8-Q4T} & \\
& Wide-field calcium imaging (ketamine) & EBRAINS Knowledge Graph & \doi{10.25493/QFZK-FXS} & \\
& Wide-field calcium imaging (isoflurane) & EBRAINS Knowledge Graph & \doi{10.25493/XJR8-QCA} & \\ \hline
& Experimental models: Organisms/strains  &  &  &  \\ \hline
& Mus Musculus: C57BL/6J & IDIBAPS & RRID:IMSR\_JAX:000664 & \\
& Mus Musculus: Del(5Gtf2i-Fkbp6)1Vcam/Vcam (WBS-KO) & IDIBAPS & & \\
& Mus Musculus: ATJ/FVB.129P2-FMR1-mix (FXS-KO)& IDIBAPS & & \\
& Mus Musculus: C57BL/6J-Tg(Thy1-GCaMP6f)GP5.17Dkim/J & LENS & RRID:IMSR\_JAX:025393 & \\ \hline
& Software and algorithms  &  &  & \\ \hline
& Analysis and plotting scripts & This paper & \url{https://github.com/INM-6/slow_wave_analysis} & \\
& Collaborative Brain Wave Analysis Pipeline (Cobrawap) & \href{https://github.com/INM-6/cobrawap}{This paper} & RRID:SCR\_022966 & \\
& Snakemake & \citet{Molder2021_33} & RRID:SCR\_003475 & \\
& Neo & \citet{Garcia2014_10} & RRID:SCR\_000634 & \\
& Nix & \citet{Stoewer2014_} & RRID:SCR\_016196 & \\
& Elephant & \citet{Denker2018_a} & RRID:SCR\_003833 & \\
& SciPy & \citet{Virtanen2020_261} & RRID:SCR\_008058 & \\ 
& Pandas & \citet{McKinney2010_56} & RRID:SCR\_018214 & \\
\bottomrule
\end{tabular}
\end{fullwidth}
\end{table}

\section{Acknowledgments}
This study was carried out in the framework of the Human Brain Project (HBP\footnote{\url{https://www.humanbrainproject.eu}}) and EBRAINS\footnote{\url{https://ebrains.eu}}. We thank Andrew Davison for his continued advice and support in developing and integrating Cobrawap.

\section{Funding}
This research was funded by the European Union's Horizon 2020 Framework Programme for Research and Innovation under Specific Grant Agreements No. 785907 (HBP SGA2) and No. 945539 (HBP SGA3); the Joint Lab “Supercomputing and Modeling for the Human Brain; the Deutsche Forschungsgemeinschaft (DFG, German Research Foundation) - 368482240/GRK2416; and the Ministry of Culture and Science of the State of North Rhine-Westphalia, Germany, under NRW-network 'iBehave' grant number NW21-049.

\section{Author contributions}
\textbf{conceptualization:} MD, PP, RG, GDB; 
\textbf{methodology:} RG, GDB, CDL, PP, MD, CC, MM, AAM, FR, AM, EP; 
\textbf{software:} RG, CDL; 
\textbf{investigation:} RG, GDB, CDL, PP, MD; 
\textbf{resources:} AAM, AM, FR, FP, MS;
\textbf{writing -- original draft preparation:} RG; 
\textbf{writing -- review and editing:} all; 
\textbf{visualization:} RG; 
\textbf{supervision:} MD, PP;
\textbf{project administration:} MD, PP, SG, MM, MS;
\textbf{funding acquisition:} PP, MD, SG, FP, MS

\textbf{Author abbreviations:} RG (\textit{Robin Gutzen}), GDB (\textit{Giulia De Bonis}), CDL (\textit{Chiara De Luca}), EP (\textit{Elena Pastorelli}), CC (\textit{Cristiano Capone}), AAM (\textit{Anna Letizia Allegra Mascaro}), FR (\textit{Francesco Resta}), AM (\textit{Arnau Manasanch}), FP (\textit{Francesco Saverio Pavone}), MS (\textit{Maria V. Sanchez-Vives}), MM (\textit{Maurizio Mattia}), SG (\textit{Sonja Grün}), PP (\textit{Pier Stanislao Paolucci}), MD (\textit{Michael Denker})

\printbibliography

\if@endfloat\clearpage\processdelayedfloats\clearpage\fi

\begin{appendix}

\begin{appendixbox}\label{app:ttt}
    
\section{Videos of example calcium imaging and ECoG wave activity recordings.}
\label{sec:wave_videos}
\begin{center}
    \includegraphics[width=.85\textwidth]{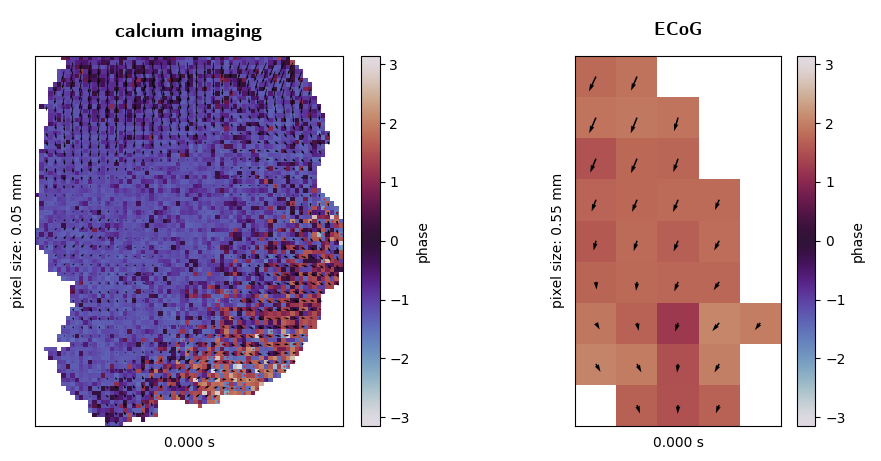}
    \captionof{figure}{\textbf{Wave videos of example calcium imaging and ECoG recordings.} The figure shows the first frames of the respective videos showing the wave activity as it is output from stage 4 of the Cobrawap pipeline. The videos are available in the online version of this paper, at \textit{\url{tbd}}.
    \label{fig:wave_videos}}
\end{center}
  
\section{Characterization of wave-modes in one ECoG recording}
\label{sec:wavemodes_dashboard_ecog}
\begin{center}
    \includegraphics[width=.85\textwidth]{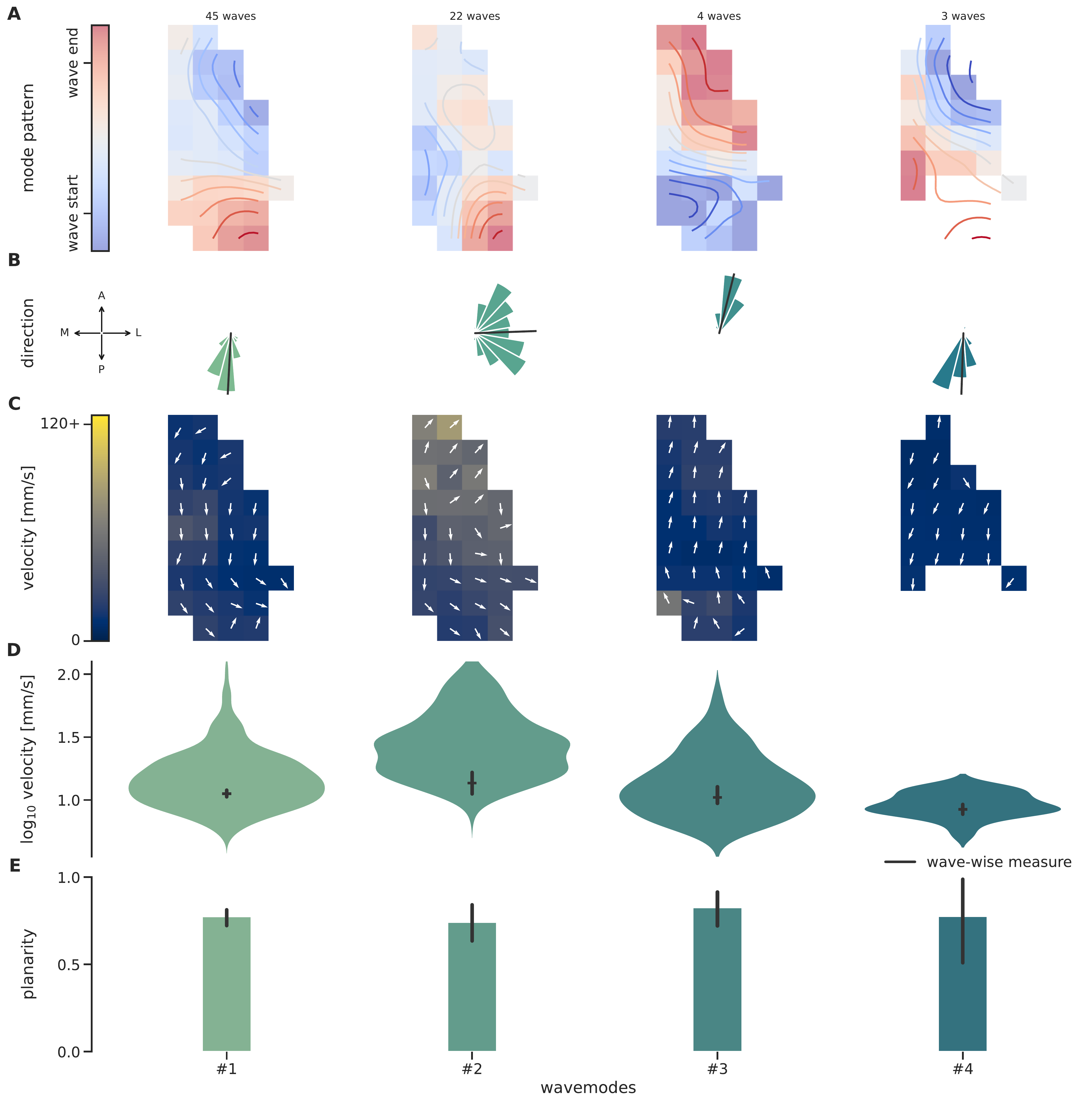}
    \captionof{figure}{\textbf{Characterization of wave-modes in one ECoG recording.} Within the pipeline, the optional block 'wave\_mode\_clustering' groups together similar wave modes. Their characterization of the waves in each of the 4 modes is shown in the corresponding columns. \textbf{A:} The average wave pattern (number of waves indicated on top) is illustrated as a time-delay heatmap with iso-delay contours. \textbf{B:} The aggregated histogram of channel-wise directions in waves of this mode. The black lines indicate the average wave-wise direction measure. \textbf{C:} Map of the average channel-wise velocities in waves of this mode, overlayed with the average channel-wise direction determined via the optical flow. \textbf{D:} The corresponding distributions of channel-wise velocities and as black ticks and errorbars the average and $95\%$ CI of the corresponding wave-wise velocities. \textbf{E:} The average and $95\%$ CI of the planarity values for the waves of this mode. The figure is analogous to \prettyref{fig:wavemodes_dashboard}.}
    \label{fig:wavemodes_dashboard_ecog}
\end{center}

\section{Data overview}
\label{sec:dataset_overview}
\begin{center}
    \includegraphics[width=.85\textwidth]{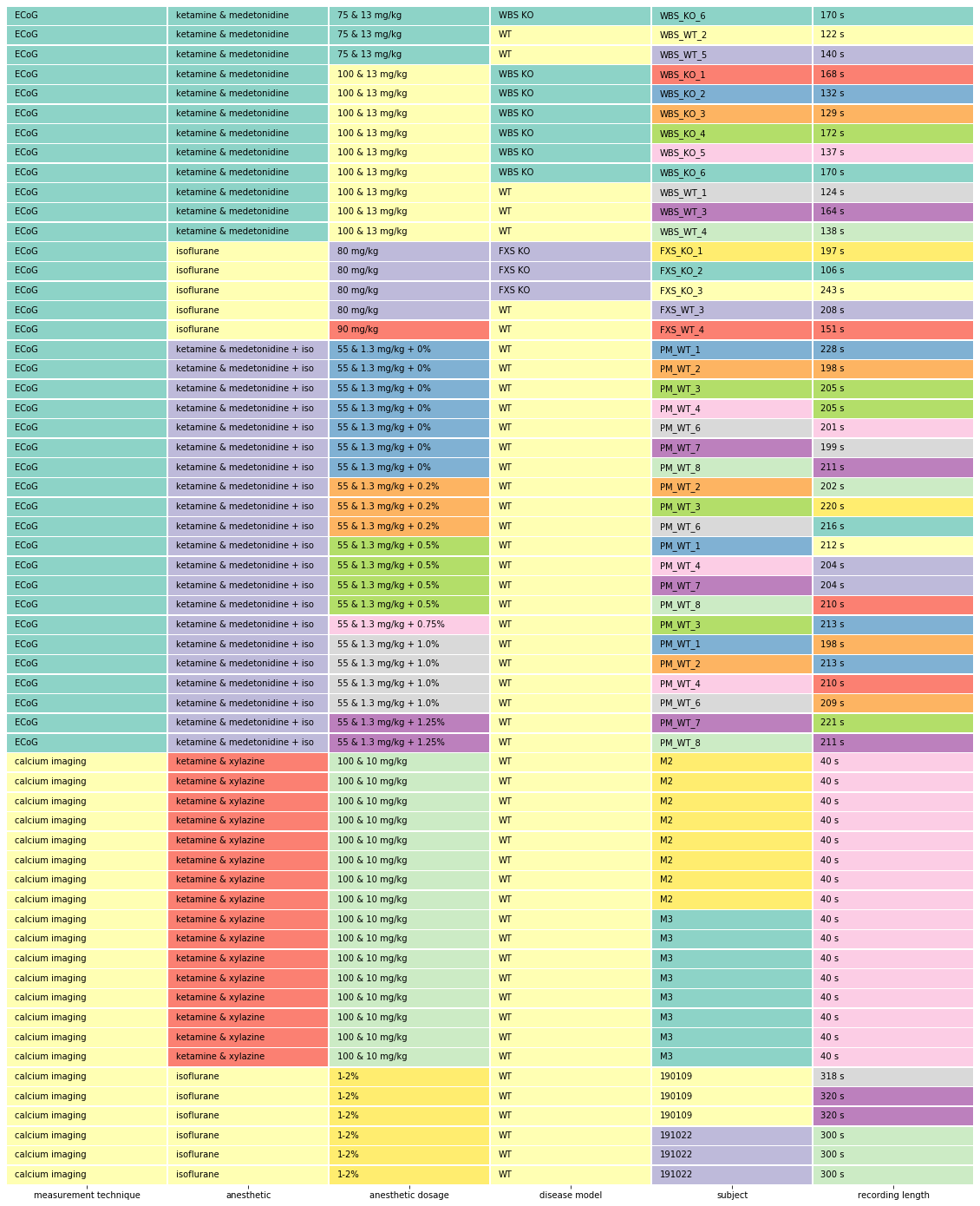}
\captionof{figure}{\textbf{Data overview.} Each row shows one of the $60$ recordings used in this study. The columns show some of the attributes in which they can differ, and within each column, different values are colored differently.
\label{fig:dataset_overview}}
\end{center}

\section{Wavefront definition via trigger clustering}
\label{sec:trigger_clustering}
\begin{center}
    \includegraphics[width=.85\textwidth]{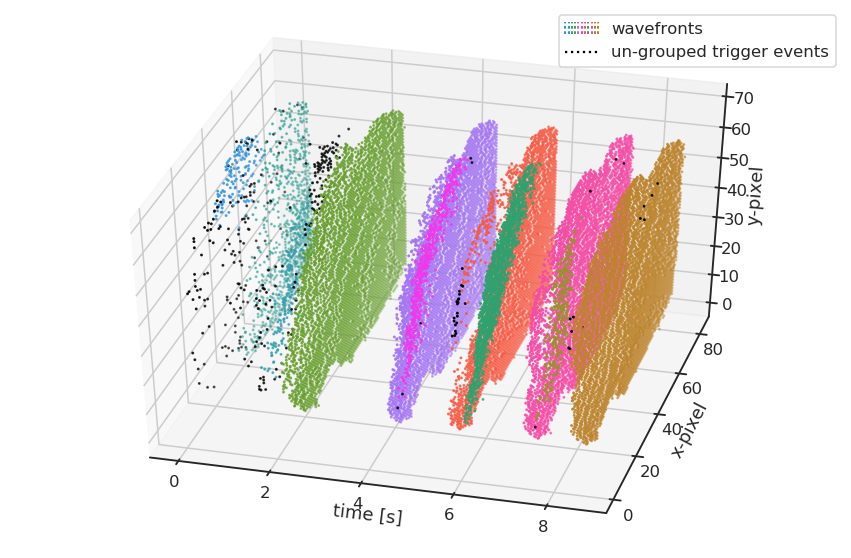}
    \captionof{figure}{\textbf{Wavefront definition via trigger clustering.} Visualizing the clustering of detected transition times in the space-time domain for $10$~s of an example calcium imaging recording. The trigger events are grouped based on their proximity in space and time using a density-based clustering algorithm (color coded).}
    \label{fig:trigger_clustering}
\end{center}
\end{appendixbox}


\end{appendix}


\end{document}